\documentclass[11pt,a4paper]{article} 
 
        \usepackage[utf8]{inputenc}
        \usepackage{fontenc}
	\usepackage[italian,english]{babel}
	\usepackage{amsmath}
	\usepackage{geometry}
	\usepackage{amsthm}
	\usepackage{mathrsfs}
	\usepackage{bbold}
	\usepackage{mathtools}
	\usepackage{amsfonts}
	\usepackage{graphicx}
	\usepackage{setspace}
	\usepackage{physics}
	 \usepackage{comment}
	 \usepackage{bbm}
	  \usepackage{layout}
	\usepackage{xcolor}
     \usepackage[big]{layaureo}
	 \usepackage{dsfont}	
	 \usepackage{slashed}
 	 \usepackage{amssymb}
	 \usepackage{tensor}
	\usepackage{fancyhdr} 
 	\usepackage{tikz-cd}
 	\usepackage{braket}
 	\usepackage{tikz}
 	\usepackage[compat=1.1.0]{tikz-feynman}
 	\usepackage{subfigure}
	\usepackage{fancyhdr} 
	\usepackage{booktabs}
	\usepackage{braket}
	\usepackage{ytableau}

\renewcommand{\arraystretch}{1.5}
	        
\usepackage{jheppub} 

\usepackage{natbib}
\title{On protected defect correlators in 3d $\mathcal{N}\ge4$ theories}

\author[a]{Luigi Guerrini}

\affiliation[a]{Dipartimento di Fisica, Universit\`a di Firenze and INFN Sezione di Firenze, via G. Sansone 1, 50019 Sesto Fiorentino, Italy}

\emailAdd{luigi.guerrini@unifi.it} 

\abstract{We study and compute supersymmetric observables for line defects in 3d $\mathcal{N}\ge4$ theories.
Our setup is a novel supersymmetric configuration involving line operators and local operators living on a linked circle. The algebra of the local operators is described by a topological quantum mechanics.
For operators belonging to conserved current multiplets, we propose an exact formula for their correlation functions based on a Ward identity for integrated correlators. Our formula gives a general recipe to compute the bremsstrahlung function for any $\frac{1}{3}$--BPS lines in $\mathcal{N}=6$ SCFTs. We apply our relation to the $\frac{1}{2}$--BPS Wilson loop in the ABJM model, showing the validity of previous computations.
Furthermore, our construction allows us to explore higher points correlators. As an example, we compute the two-point function of the stress tensor multiplet correlators in ABJM theory in the presence of the Wilson line. We also present some perturbative checks of our formulae.}

	\newcommand{\beq}{\begin{equation}}
	\newcommand{\bea}{\begin{eqnarray}}
	\newcommand{\eea}{\end{eqnarray}}
	\newcommand{\eeq}{\end{equation}}
	\newcommand{\non}{\nonumber}

	\renewcommand{\a}{\alpha}
	\renewcommand{\b}{\beta}

	\makeatletter
	\newcommand{\aextp}{\@ifnextchar^\@aextp{\@aextp^{\,}}}
	\def\@aextp^#1{\mathop{\bigwedge\nolimits^{\!#1}}}
	\makeatother
	
	\makeatletter
	\newcommand{\extp}{\@ifnextchar_\@extp{\@extp_{\,}}}
	\def\@extp_#1{\mathop{\aextp\nolimits_{\!#1}}}
	\makeatother

	\theoremstyle{definition}

\begin{document}
\maketitle
\section{Introduction}

In conformal field theories (CFTs), the fundamental observables are correlation functions of local operators. In principle, they are completely determined by the spectrum, i.e., by the dimensions of the operators and the coefficients of the operator product expansion (OPE). Together they form the CFT data.
CFTs can also be enriched by extended defects that preserve the conformal symmetry of the worldvolume. These are conformal defects, and relevant examples are magnetic impurities and line operators in conformal gauge theories.
Because of the reduced symmetry, more observables are available: These include the spectrum of defect operators and their defect OPE (operator product expansion) and the bulk-to-boundary OPE coefficients \cite{Billo:2016cpy}.
Despite the remarkable progress in recent years based on the bootstrap approach \cite{Rattazzi:2008pe}, computing CFT data with or without conformal defects is still a highly non-trivial task.

Superconformal field theories (SCFTs) are distinguished models where some CFT data can be computed exactly. For instance, the correlation functions of operators preserving a fraction of supersymmetry, called BPS operators, can be protected from quantum corrections. That is the case for the two and three-point functions of the $\frac{1}{2}$--BPS local operators in $\mathcal{N}=4$ SYM \cite{Lee:1998bxa}. Building on this example, the authors of \cite{Drukker:2009sf} found a class of local operators preserving the same supersymmetries regardless of the number of operator insertions.
The crucial insight is to endow the operators with a controlled spacetime dependence via a conformal twist \cite{deMedeiros:2001wqm}.
This idea has been recently reinvestigated in \cite{Beem:2013sza} from the superconformal bootstrap point of view and extended to $\mathcal{N}=2$ SCFTs. Remarkably, a sector of the full operator algebra is described by a chiral algebra provided that operators lie in the same plane (see also \cite{Beem:2014kka} for the analogous construction in 6d).
Similarly, in 3d $\mathcal{N}\ge4$ SCFTs correlation functions of a subset of operators placed on a line are captured by a topological field theory \cite{Chester:2014mea, Beem:2016cbd, Liendo:2015cgi}. We call this set the \emph{topological sector}, and it will be one of the main characters of this work.

One of the advantages of the protected subsectors is the possibility of using supersymmetric localization to access some of the associated CFT data.
For example, a localization scheme for the topological sector in UV Yang-Mills theories flowing to interacting $\mathcal{N}\ge4$ SCFTs
was proposed in \cite{Dedushenko:2016jxl, Dedushenko:2017avn, Dedushenko:2018icp} for operators on $S^3$ and in \cite{Panerai:2020boq} for operators on $S^2\times S^1$. However, a complete localization is not always accessible. Alternatively, one can relate specific operators of the SCFT to exactly computable supersymmetric deformations of the theory. For example, Coulomb branch operators in 4d $\mathcal{N}=2$ SCFTs are related to marginal deformations of the models \cite{Gerchkovitz:2016gxx}. Similarly, in 3d $\mathcal{N}\ge4$ SCFTs, certain integrated topological operators are captured by supersymmetric mass deformations of the partition function \cite{Agmon:2017xes, Binder:2019mpb, Binder:2020ckj, Gorini:2020new, Guerrini:2021zuk, Bomans:2021ldw}. This strategy is particularly convenient when the deformed partition function is known exactly.

A natural generalization is to study superconformal field theories with superconformal defects. They are examples of tractable conformal defects, which can be studied with a variety of techniques, like localization, bootstrap, integrability, and holography. Following the same logic of the previous examples, supersymmetric setups with both non-local and local operators would provide us with a powerful tool to investigate the physics of the defect.
Relevant examples in $\mathcal{N}=4$ SYM involve the $\frac{1}{2}$--BPS defects, which can be the BPS Wilson line \cite{Maldacena:1998im}, surface defects \cite{Gukov:2006jk}, and boundaries/interfaces \cite{Gaiotto:2008sa, Gaiotto:2008ak} together with the protected local operators mentioned above \cite{Drukker:2009sf}. In these cases, localization schemes are available \cite{Giombi:2009ds, Wang:2020seq}.
Less supersymmetric examples in 4d involve Coulomb branch operators and $\frac{1}{2}$--BPS Wilson loops in $\mathcal{N}=2$ gauge theories \cite{Billo:2018oog}, and chiral algebra and surface defects in 4d $\mathcal{N}=2$ SCFTs \cite{Bianchi:2019sxz, Pan:2017zie}. 

This paper aims to explore the corresponding problem for 3d SCFTs with BPS line operators, such as Wilson and vortex loops \cite{Gaiotto:2007qi, Drukker:2008jm, Chen:2008bp, Rey:2008bh, Drukker:2009hy, Drukker:2008zx, Kapustin:2012iw, Drukker:2012sr}. These are interesting extended operators with holographic duals and exhibit non-trivial mapping under IR dualities 
\cite{Assel:2015oxa, Dimofte:2019zzj, Dey:2021jbf, Dey:2021gbi, Griguolo:2021rke, Thull:2022lif}. Since in 3d the only known family of non-trivial protected local operators live in $\mathcal{N}=4$ theories, we restrict to these models. 

\begin{figure}[h]\label{flat:conf}
	\centering
\includegraphics[scale=0.65]{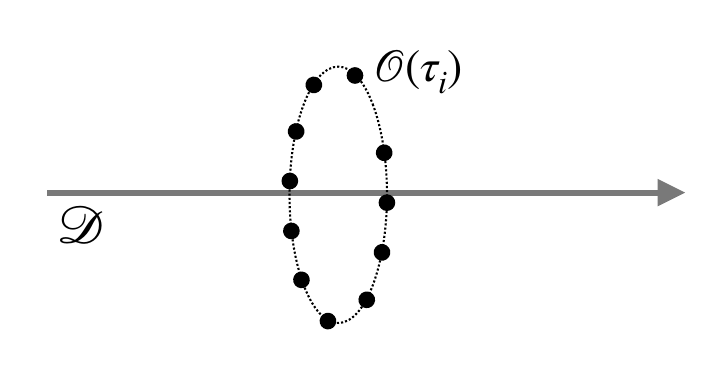}
\caption{We sketch the kinematics of the flat space configuration. The line defect $\mathcal{D}$ extends along a straight line, while the local operators $\mathcal{O}(\tau_i)$ live on the circle in the orthogonal plane to the defect.}
\end{figure}

To fix ideas, we begin to study theories with a UV Lagrangian description. Having in mind localization, we work directly on $S^3$. These theories admit $\frac{1}{2}$--BPS Wilson and vortex loops on any great circle. Even if they are not invariant under conformal symmetry, they are believed to flow to non-trivial conformal defects \cite{Assel:2015oxa}. 
From a detailed analysis of the preserved supersymmetry algebras, we find a supersymmetric configuration with a topological sector on the great circle linking with the loop operator.
At the CFT point, the corresponding flat space setup has the defect extended on a straight line, and the topological operators on a circle in the orthogonal plane, as sketched in figure~\ref{flat:conf}.

We emphasize that unlike previous studies that focused on defect operators living on the extended defect \cite{Dedushenko:2016jxl, Dimofte:2019zzj}, our configuration involves genuine local operators away from the loop. The two configurations preserve different supercharges. Moreover, at the CFT point, the two setups compute different observables. On the one hand, correlation functions with only defect operators compute the defect OPE. On the other hand, our system allows us to access the bulk-to-boundary OPE. These two sets of data are related by crossing symmetry \cite{Liendo:2012hy}.

More precisely, for any BPS Wilson loop, we identify a Coulomb branch topological sector, i.e. a set of operators built out of the vector multiplet.
We describe also a configuration involving vortex loops and the Higgs branch topological sector, whose fields are made by scalars in the hypermultiplet. Based on the algebra preserved by the supercharges and the local nature of the duality operation \cite{Gaiotto:2008ak, Gulotta:2011si, Hwang:2021ulb}, we expect these two configurations to be related by mirror symmetry. We then extend these examples to any $\frac{1}{2}$--BPS superconformal line operators in a generic $\mathcal{N}=4$ SCFT.

Then, we show how to exploit these configurations to extract CFT data. The main tools are localization and the Ward identity relating mass deformations and dimension one topological operators. Using the Ward identity, we argue that the vev of the defect operator in a properly mass-deformed background is the generating functional for integrated defect correlation functions, generalizing the argument without defects. Using localization, one can sometimes reduce the computation of the vev to a finite-dimensional integral, i.e. a matrix model, which can be evaluated in different limits.

As concrete applications, we evaluate the correlation functions with dimension one topological operators and the $\frac{1}{2}$--BPS Wilson line in the $U(N_1)_k\times U(N_2)_{-k}$ ABJM model.
The ABJM model is $\mathcal{N }= 6$ Super Chern-Simons-matter theories, dual to type IIA string theory in AdS$_4\times \mathbb{CP}_3$ or M-theory in AdS$_4\times S^7/Z_k$ \cite{Aharony:2008ug, Aharony:2008gk}.
The relevant operator is a specific bilinear of the matter scalar fields, which is also part of the stress tensor multiplet. We evaluate the corresponding one and two-point functions by taking mass derivatives of the mass-deformed vev of the Wilson loop, which is known from localization \cite{Kapustin:2009kz, Jafferis:2010un, Hama:2010av}.

According to \cite{Lewkowycz:2013laa}, the one-point function of the stress tensor is proportional to the so-called bremsstrahlung function \cite{Correa:2012at}. 
Being this quantity often accessible to integrability, it provides a potentially fruitful bridge between different methods.
We expand our matrix model prediction for large values of the Chern-Simons coupling $k$. In this limit, the theory becomes perturbative, and we compare our results with those obtained from standard Feynman diagrams. We perform this check explicitly at order $1/k$. Another reason to consider this limit is the abundance of results for the bremsstrahlung function. We find perfect agreement with all the data available in the literature \cite{Griguolo:2012iq, Lewkowycz:2013laa, Bianchi:2014laa, Bianchi:2017svd, Bianchi:2017ozk, Bianchi:2018bke, Bianchi:2018scb, Griguolo:2021rke}. In the case $N_1 \neq N_2$, this provides the first complete derivation of the bremsstrahlung, confirming the conjecture of \cite{Bianchi:2018bke}. The comparisons also fix the proportionality constant between the bremsstrahlung and the derivatives. Since this constant is universal, we propose a formula for the bremsstrahlung valid for any $\frac{1}{3}$--BPS line defect \cite{Drukker:2022txy}.

As far as we know, the result for the two-point function is completely new. We consider that because it is the simplest observable exhibiting crossing symmetry between the bulk OPE and the bulk-to-boundary OPE. Thus, one could use our result to extract defect CFT data, which might be accessible using other methods such as bootstrap and, perhaps, integrability. Our computation provides a potential cross-check for those techniques. Even in this case, we compare the result with the one obtained from standard perturbative methods.

The paper is organized as follows. In Sec.~\ref{sec1} we introduce the theories and the line operators we will study. Then, in Sec.~\ref{sec2}, after a brief review of the topological sector, we combine it with local operators, both for UV gauge theories and SCFTs. In Sec.~\ref{sec3} we present our formula relating integrated defect correlation functions to mass deformations, and we apply it to ABJM. Finally, in Sec.~\ref{sec4} we discuss our results and compare against perturbative computations.
 We end the paper with a general discussion on the results and possible future directions. Technical details and conventions are provided in two appendices.

\section{Line operators in $\mathcal{N}\ge 4$ theories}\label{sec1}

In this section, we give an overview of BPS line operators in 3d $\mathcal{N}\ge 4$ gauge theories, both with and without Chern-Simons terms. Then, we discuss what observables are computable for conformal defects, with some focus on those related to the stress tensor.

\subsection{Loop operators in UV gauge theories}\label{sec1.1}

To begin with, we discuss gauge theories without Chern-Simons terms \cite{Dedushenko:2016jxl}. Since we aim to make contact with localization, we find it convenient to define these QFTs directly on the three-sphere $S^3$. 
We work in toroidal coordinates $\theta\in[0,\frac{\pi}{2})$, $\varphi\in[0,2\pi)$, $\tau\in[0,2\pi)$. The metric reads
\begin{equation}\label{eq:metric}
ds^2=r^2\left(d\theta^2+\sin^2\theta \,d\varphi^2+\cos^2\theta\, d\tau^2\right)\,.
\end{equation}
The name toroidal is because at fixed $\theta$ the metric is a 2d torus. When $\theta=0$, the metric shrinks to a circle $S^1_\varphi$, while at $\theta=\pi/2$, the metric shrinks to a circle $S^1_\tau$.
The subscripts $\varphi,\, \tau$ indicate the variables parametrizing the circle in these coordinates \footnote{More details on the geometry of $S^3$ can be found in \cite{Dedushenko:2016jxl}.}.
These two great circles will play a central role in our construction

The building blocks are vector multiplets and hypermultiplets.
The $\mathcal{N}=4$ vector multiplet $\mathcal{V}$ contains the gauge vector $A_\mu$, the gaugino $\lambda_{\alpha,a\dot{a}}$, the dimension one scalar $\Phi_{\dot{a}\dot{b}}$, and the auxiliary field $D_{ab}$, which has dimension 2. They transform respectively in the singlet, $(\mathbf{2},\mathbf{2})$,  $(\mathbf{3},\mathbf{1})$, and $(\mathbf{1},\mathbf{3})$ representation of the R-symmetry group $SU(2)_C\times SU(2)_H$. All the components transform in the adjoint representations of the gauge group $G$. 
The matter is organized in hypermultiplets $\mathcal{H}$, whose field components are the $SU(2)_H$ doublets $q_a$ and $\tilde{q}^a$, and the fermions $\psi_{\dot{a}}$ and $\tilde{\psi}^{\dot{a}}$, which are doublets of $SU(2)_C$. $q_a$ and $\psi_{\dot{a}}$ transform in the unitary representation $\mathcal{R}$ of $G$, $\tilde{q}^a$ and $\tilde{\psi}^{\dot{a}}$ in the complex conjugate representation $\bar{\mathcal{R}}$.

The action for the hypers can be obtained by conformally mapping the flat space one on $S^3$. We get
\begin{align}\label{eq:Shyper}
S_{\textup{hyper}}&=\int d^3x\sqrt{g}\Bigg[ \mathcal{D}^\mu\tilde{q}^a\mathcal{D} _\mu q_a - i\tilde{\psi}^{\dot{a}}\gamma^\mu\mathcal{D}_\mu\psi_{\dot{a}}+\frac{3}{4r^2}\tilde{q}^a q_a+i\tilde{q}^a{D_a}^b q_b-
\frac{1}{2}\tilde{q}^a\Phi^{\dot{a}\dot{b}}\Phi_{\dot{a}\dot{b}} q_a+\notag\\
&-i\tilde{\psi}^{\dot{a}}{\Phi_{\dot{a}}}^{\dot{b}}\psi_{\dot{b}}+i\left(\tilde{q}^a{\lambda_a}^{\dot{b}}\psi_{\dot{b}}+\tilde{\psi}^{\dot{a}}{\lambda^{b}}_{\dot{a}} q_b\right)\Bigg] \,.
\end{align}
The action is invariant under the variations \eqref{susyh}.
If we think of the gauge field as a background field, this action is conformally invariant.
Writing down a superconformal action for the \emph{off-shell} vector multiplet is a hard task \cite{Kuzenko:2015lfa}, and we do not pursue it here. However, it is possible to give up conformal invariance and write down a supersymmetric Yang-Mills term 
\begin{equation}
\begin{aligned}
\label{eq:Yang_Mills_action}S_{\textup{YM}}=&\frac{1}{g^2_{\mathrm{YM}}}\int d^3x\sqrt{g}\Tr\bigg[F^{\mu\nu}F_{\mu\nu}-\mathcal{D}^\mu\Phi^{\dot{a}\dot{b}}\mathcal{D}_\mu\Phi_{\dot{a}\dot{b}}+i\lambda^{a\dot{a}}\gamma^\mu\mathcal{D}_\mu \lambda_{a\dot{a}}-D^{ab}D_{ab}+\\
&-i\lambda^{a\dot{a}}\comm*{{\lambda_a}^{\dot{b}}}{\Phi_{\dot{a}\dot{b}}}-\frac{1}{4}\comm*{{\Phi^{\dot{a}}}_{\dot{b}}}{{\Phi^{\dot{c}}}_{\dot{d}}}\comm*{{\Phi^{\dot{b}}}_{\dot{a}}}{{\Phi^{\dot{d}}}_{\dot{c}}}-\frac{1}{2r}h^{ab}\bar{h}^{\dot{a}\dot{b}}\lambda_{a\dot{a}}\lambda_{b\dot{b}}+\\
&+\frac{1}{r}\left({h_a}^b{D_b}^a\right)\Bigl({\bar{h}^{\dot{a}}}_{\dot{b}}{\Phi^{\dot{b}}}_{\dot{a}}\Bigr)-\frac{1}{r^2}\Phi^{\dot{a}\dot{b}}\Phi_{\dot{a}\dot{b}}\bigg] \,.
\end{aligned}
\end{equation}
This action is invariant under the variations \eqref{susyN4}, when restricted to the Poincar\'e subalgebra $\mathfrak{su}(2|1)_\ell\oplus\mathfrak{su}(2|1)_r$. Its bosonic part contains the isometry group of the sphere $SO(4)\simeq SU(2)_\ell\times SU(2)_r$, and the Cartan of the $SO(4)$ R-symmetry group $U(1)_\ell\times U(1)_r$.
An essential ingredient in its definition is the matrices ${h_a}^b\in\mathfrak{su}(2)_H$, ${\bar{h}_{\dot a}}^{\;\;\dot{b}}\in\mathfrak{su}(2)_C$ which reduces the original R-symmetry to the Cartan \footnote{For explicit calculations we choose ${h_a}^b=-{(\sigma_2)_a}^b$, ${\bar h_{\dot a}}^{\;\;\dot b}=-{(\sigma_3)_{\dot a}}^{\;\;\dot b}$.}. 
The superalgebra has eight fermionic generators $Q_\alpha^{(\ell\pm)}$, $Q_\alpha^{(r\pm)}$, closing respectively on the left and right part of the bosonic algebra. The relevant details and conventions are summarized in the App.~\ref{appb}.
In the end, the theory preserves eight supercharges. We postpone the discussion on real mass and FI deformations to section \ref{sec3}, where they will play a central role to perform computations. 

Gauge theories contain various loop operators. Although theories with extended supersymmetry may admit loop operators on multiple closed paths, we shall focus on BPS Wilson loops supported on great circles, e.g. $S^1_\varphi$. The first example is perhaps the best known, namely the BPS Wilson loops whose generic expression is obtained by adding to the standard Wilson loop a coupling, parametrized by an arbitrary symmetric matrix $M_{\textup{Wilson}}^{\dot a\dot b}$, to the scalar $\Phi_{\dot a\dot b }$ in the vector multiplet
\begin{equation}
W_R[\gamma]=\Tr \,P\exp\left[ i\int_{\gamma} ds\left(\dot{x}^\mu A_\mu+ir\Phi_{\dot a\dot b}M_{\textup{Wilson}}^{\dot a\dot b}\right)  \right]\,,
\end{equation}
where $R$ indicates a representation of the gauge group.
Restricting the curve $\gamma$ to be $S^1_\varphi$ and requiring SUSY, we get a $\frac{1}{2}$--BPS Wilson loop, whose explicit form is
\begin{equation}\label{uvwl}
W_R=\Tr \,P\exp\left[ i\int_{S^1_\varphi} d\varphi(A_{\varphi}-ir\Phi_{\dot 1\dot 2})  \right]\,.
\end{equation}
It preserves $Q^{\ell+}_2$, $Q^{(\ell-)}_1$, $Q^{(r+)}_2$, and $Q^{(r-)}_1$, generating a $\mathfrak{su}(1|1)\oplus\mathfrak{su}(1|1)$ superalgebra.

Another distinguished class of line operators is vortex loops.
They are defect operators defined by singular classical BPS configurations for the vector field. The field strength $F_{\mu\nu}$ is proportional to a delta function supported on the loop worldvolume. Their BPS version was introduced in \cite{Kapustin:2012iw, Drukker:2012sr} and is realized by turning on a singular imaginary profile for the auxiliary field. 
In the $\mathcal{N}=4$ language, the form of an abelian BPS vortex loop of charge $q$ reads
\begin{equation}\label{vl}
F=2\pi q\delta_{\ell},\quad D^{ab}=-2\pi q\star\left(\delta_{\ell}\wedge d\ell\right)M_{\text{vortex}}^{ab}\,,
\end{equation}
where $d\ell$ is the one form dual to $\dot{\ell}$, and $\delta_{\ell}$ is the Poincar\'{e} dual to the loop worldvolume.
The matrix $M_{\text{vortex}}^{ab}$ is chosen in combination with the curve $\gamma$ to set to zero the variation of the gaugino for some specific supercharge. Again, we shall restrict to operators supported on $S^1_{\varphi}$. If we also choose $M_{\text{vortex}}^{ab}=-\delta^{ab}$, the vortex loop preserves half of the supersymmetry, namely $Q^{(\ell+)}_1$, $Q^{(\ell-)}_2$, $Q^{(r+)}_2$, and $Q^{(r-)}_1$ generating another $\mathfrak{su}(1|1)\oplus\mathfrak{su}(1|1)$ superalgebra.
To get non-abelian vortex loops, one replaces the number $q$ with an element of the Lie algebra $\mathfrak{g}$\footnote{An alternative definition is proposed in \cite{Assel:2015oxa}. The vortex loop is represented by adding one-dimensional local degrees of freedom on the curve supporting the loop operators. Integrating out the 1d d.o.f. reproduces the vortex singularity for the bulk fields. See also \cite{Hosomichi:2021gxe} for a recent discussion.}.  

\subsection{BPS lines in Chern-Simons matter theory (ABJM)}\label{sec1.2}

Including a Chern-Simons term preserving $\mathcal{N}\ge 4$ supersymmetry requires some more work. As far as we know, no formulation with off-shell vector multiplets preserving the full superconformal algebra is known.
Nevertheless, Chern-Simons theories with $\mathcal{N}\ge 4$ supersymmetry arise from specific lower supersymmetric theories after integrating out of all the auxiliary fields. This idea was introduced in \cite{Gaiotto:2008sd} and further generalized in \cite{Hosomichi:2008jd, Imamura:2008dt}. We will refer to these theories as generalized Gaiotto-Witten theories. 
The ABJM model, which is a $\mathcal{N}=6$ quiver theory with gauge group $U(N_1)_k\times U(N_2)_{-k}$ (see App.~\ref{appa} for more details), falls into this category \cite{Aharony:2008ug, Aharony:2008gk, Hosomichi:2008jb}.
Generalized Gaiotto-Witten theories contain many interesting BPS loop operators \cite{Cooke:2015ila}. However, since our applications will be in ABJM, we explicitly describe them only in this case.
There are no conceptual difficulties to extend our construction to those more general Chern-Simons theories.

The construction of a maximally supersymmetric Wilson loop in ABJM is somewhat tricky and requires to think of the gauge group $U(N_1)\times U(N_2)$ embedded in the supergroup of $U(N_1|N_2)$. Then, we consider the holonomies of this larger group \cite{Drukker:2009hy}. That is, we take line operators of the form
\begin{equation}\label{WL}
W_\mathcal{R}[C] =\Tr_\mathcal{R}P\exp (-i\int_C \mathcal L)\,,
\end{equation}
where $\mathcal{R}$ is a representation of $U(N_1|N_2)$.
For a generic smooth path $C$ in ${\mathbb R}^3$ parametrized as $x^\mu=x^\mu(\tau)$, the general structure of $\mathcal{L}$ is given by
\begin{equation}\label{connec}
\mathcal L=
\begin{pmatrix}
 A_\mu\dot x^\mu-  \frac{2\pi i}{k}|\dot x|{M_J}^IC_I\bar C^J & i\sqrt{2\pi\frac{\ell}{k}}|\dot x|\eta^\alpha_I\bar\psi^I_\alpha\\
-i\sqrt{2\pi \frac{\ell}{k}}|\dot x|\psi_I^\alpha\bar\eta_\alpha^I& \hat A_\mu\dot x^\mu-  \frac{2\pi i}{k}|\dot x|{\hat M_J}^I\bar C^JC_I
\end{pmatrix}\,.
\end{equation}
Sometimes we refer to the resulting operator as the ``fermionic'' Wilson loop because matter fermions appear naturally in the off-diagonal entries of $\mathcal L$.
The possible local quantities $M_{J}^{\ \ I}(\tau)$, $\hat M_{J}^{\ \ I}(\tau)$, $\eta_{I}^{\alpha}(\tau)$ and $\bar{\eta}^{I}_{\alpha}(\tau)$ regulate the coupling of the matter fields to the loop.
Imposing local $U(1)\times SU(3)$ $R-$symmetry and local supersymmetry invariance, we find a general expression for the couplings in \eqref{connec} 
\begin{align}
\eta_{I}^{\alpha} (\tau)&=n_{I} (\tau)\eta^{\alpha} (\tau)\,,\qquad\quad  \bar\eta^{I}_{\alpha} (\tau)=\bar n^{I} (\tau) \bar\eta_{\alpha} (\tau), \notag\\
M_{J}^{\ \ I} (\tau)&=
\widehat M_{J}^{\ \ I} (\tau)=\delta^{I}_{J}-2 n_{J} (\tau) \bar n^{I} (\tau)\,.
\end{align}
Following \cite{Drukker:2009hy, Cardinali:2012ru, Lietti:2017gtc}, one can determine the functions of $n^I(\tau)$ and $\eta_\alpha(\tau)$ on $\tau$ by imposing invariance under some superconformal transformations. Crucially, we demand a more general BPS condition w.r.t. to Wilson loop in Yang-Mills theories. Unlike the case of UV gauge theories, where we require that a supersymmetry variation annihilates the connection of the Wilson loop in \eqref{uvwl}, in Chern-Simons matter theories we only impose the variation to be a supergauge transformation
\begin{equation}
\delta W_\mathcal{R}[C] =\comm{G(\tau)}{W_\mathcal{R}[C] }\,,
\end{equation} 
where $G(\tau)$ is a field-dependent super-gauge transformation belonging to $U(N_1|N_2)$. This weaker BPS condition makes the Wilson loop invariant only after taking the trace, forcing us to look carefully at the form of $G(\tau)$. Since for closed path $G(\tau)$ is not periodic, we need to modify the definition \eqref{WL} by adding a twist matrix $\mathcal{T}$ restoring the correct periodicity of $G(\tau)$. In the end, the fermionic Wilson loop is
\begin{equation}
\mathcal{W}_{\mathcal{R}}[C] =\mathrm{Str}\left( W_\mathcal{R}[C]  \mathcal{T}\right)\,.
\end{equation}
See \cite{Cardinali:2012ru} for details\footnote{Alternatively, this issue can be cured by introducing a classical background connection along the path, which makes the super-gauge transformation periodic \cite{Drukker:2019bev}.}.
When the path $C$ is the infinite straight line parametrized by $x_3$, i.e. $x^\mu = (0,0, s)$, $ -\infty < s < +\infty$, and with the following choices for the couplings
\begin{equation}\label{eq:couplings}
{M_J}^I={\hat M_J}^I=
\begin{pmatrix}
-1&0&0&0\\
0&1&0&0\\
0&0&1&0\\
0&0&0&1
\end{pmatrix},
\qquad \eta^\alpha_I=\sqrt{2}
\begin{pmatrix}
1\\
0\\
0\\
0
\end{pmatrix}_I
(1,0)^\alpha,\qquad
\bar\eta_\alpha^I=i\sqrt{2}\ (1,0,0,0)^I
\begin{pmatrix}
1\\
0
\end{pmatrix}_\alpha\,,
\end{equation}
the operator preserves half of the supersymmetry charges \cite{Cardinali:2012ru}, {\em i.e.} it defines a $1/2$ BPS linear defect. See App.~\ref{appb} for details on the preserved superalgebra.
Even if it is not strictly required by supersymmetry, we introduce in the definition a twist matrix $\sigma_3$ in the trace (see \cite{Drukker:2009hy,Cardinali:2012ru}).
The reason for this is that we want to be able to map the Wilson line operator conformally to the circular Wilson loop on $S^3$ in order to perform computations consistently. Since the fermionic Wilson loop on the great circle of $S^3$ requires the twist matrix, we insert it for the Wilson line as well. This choice amounts to taking the trace instead of the super-trace of \eqref{WL}, and it is also the prescription that leads to an operator that is dual to the 1/2 BPS string configuration in AdS$_3 \times {\mathbb {CP}}^3$ or a 1/2 BPS M2-brane configuration in M-theory \cite{Lietti:2017gtc}. To simplify the notation, from now on, we will limit to denote the $\frac{1}{2}$--BPS Wilson loop as $W$. We will comment when the omitted details become relevant.

\subsection{Line operators as conformal defects}\label{sec1.3}

We review some properties of conformal defects \cite{Billo:2016cpy}. By definition, a conformal line breaks the 3d conformal symmetry down to $SO(2)\times SL(2,\mathbb{R})$. Therefore, $\frac{1}{2}$--BPS Wilson line in ABJM is an example of a conformal line defect. Sometimes, it is also possible to engineer a UV avatar for a conformal line. That is the case for the loops described in Sec.~\ref{sec1.1}. Even though the loop operators in the UV theory do not preserve conformal symmetry, there is strong evidence from localization and dualities that they flow to non-trivial conformal defects in the deep IR. 
We will briefly review some relevant features of the physics of conformal line defects.
The main point will be that the knowledge of the dimensions of all bulk operators and all their three-point functions no longer exhausts the possible CFT data.
Even though all our examples are Lagrangian, the following discussion depend on it at all. 

For simplicity, we limit to straight lines in flat three-dimensional space.
We split the coordinates $x^\mu=(x^i,s)$, where $x^i$ indicates the transverse coordinates $i=1,2$ and $s$ the coordinates along the defects. The defect $\mathcal{D}$ is placed at the position $x_i=0$.
We introduce the following notation for defect correlation functions
\begin{equation} 
\langle O_1(x_1^{i_1})\dots O_n(x_n^{i_n})\rangle_{\mathcal{D}}=\frac{1}{\langle \mathcal{D}\rangle}\langle O_1(x_1^{i_1})\dots O_n(x_n^{i_n})\mathcal{D}\rangle\,,
\end{equation}
where $\langle \mathcal{D}\rangle$ is the vev of the defect. 
The first novelty is that the defect one-point functions can be non-zero. For scalar operators, it is not hard to show that
\begin{equation}
\langle O(x) \rangle_{\mathcal{D}}=\frac{h_O}{|x^i|^{\Delta_O}}\,,
\end{equation} 
where $\Delta_O$ is the dimension of $O$ and $|x^i|$ its distance from the defect. Therefore, $h_O$ are new dynamical data of the CFT with the defect. 

A particularly relevant example is the defect one-point function of the stress tensor $T_{\mu\nu}$. One can show that (see e.g. \cite{Kapustin:2005py})
\begin{equation}\label{ht}
\langle T^{\mu\nu}(x) \rangle_{\mathcal{D}}=\frac{h_{T}}{|x^i|^d}H^{\mu\nu}\,.
\end{equation}
$H^{\mu\nu}$ is a symmetric traceless tensor 
\begin{equation}
H^{ij}=-\left(\frac{2}{3}\delta^{ij}-\frac{x^ix^j}{|x^i|^2}\right)
\,,\quad H^{ab}=\frac{1}{3}\delta^{ab}\,, \quad H^{ia}=0\,.
\end{equation}
Since the stress tensor is always defined in a local CFT, $h_T$ is a universal observable defined in any CFT. Higher spinning operators can have a non-vanishing one-point function if one succeeds in forming a tensor in the appropriate representation. However, their discussion is beyond our purposes, and we refer the reader to \cite{Billo:2016cpy}.

The second novelty is the possibility of having defect excitations, described by operators $\hat O(s)$ living on the defects. Their correlation functions, denoted as
\begin{equation} 
\langle \hat  O_1(s_1)\dots \hat O_k(s_k)\rangle_{\mathcal{D}}=\frac{1}{\langle \mathcal{D}\rangle}\langle \hat  O_1(s_1)\dots \hat O_k(s_k)\mathcal{D}\rangle\,,
\end{equation}
are constrained in the usual way by the defect conformal group $SL(2,\mathbb{R})$. Therefore, they form a peculiar conformal theory called the defect conformal field theory (dCFT).
For example, for a conformal Wilson line $W$ in a Lagrangian theory, we can give an explicit expression for the corresponding dCFT. The operators $\hat O_i(s)$ are combinations of fundamental fields transforming in the same representation of the gauge group as the Wilson loop's connection. If we denote as $W_{s_i,s_j}$ the untraced Wilson line segment starting from $s_i$ and ending on $s_j$, we can write a rather explicit expression for correlation functions
\begin{equation} 
\langle \hat  O_1(s_1)\dots \hat O_k(s_k)\rangle_{W}=\frac{1}{\langle W\rangle}\langle \Tr( W_{-\infty,s_1}\hat  O_1(s_1)\dots W_{s_{k-1},s_k} \hat O_k(s_k)W_{s_k,\infty})\rangle\,.
\end{equation} 
  
The dCFT satisfies the axioms of a standard CFT. Thus, correlators of the dCFT will be determined by the defect spectrum $\Delta_{\hat O}$ and by the coefficients of the defect three-point functions $\hat c_{\hat O_i,\hat O_j,\hat O_k}$.
 
A relevant example of a defect operator is the \emph{displacement operator}. The insertion of the defect modifies the Ward identities for broken symmetries by contact terms living localized on the worldline. The displacement operator $\mathsf{D}^i$ is responsible for such contributions for the broken transverse translations $P^i$. Explicitly, one can show that 
\begin{equation} \label{eq4.4}
 \partial_\mu T^{\mu i}(x)=-\delta_{\mathcal{D}}(x^i)\mathsf{D}^i(x^a)\,.
\end{equation}
The definition holds inside correlation functions. The physical interpretation is that the displacement operator describes the energy exchanges between the bulk and the defect.  
We observe that the displacement is defined up to primary defect operators. We fix them by requiring that $T^{\mu\nu}$ is a conformal primary near the defect. In turn, this implies that the displacement is a defect primary, whose dimension is fixed by the Ward identity to be $\Delta_{\mathsf{D}}=2$. Being the normalization of the displacement fixed by the Ward identity, the coefficient $B$ of its two-point function is a physical quantity
\begin{equation}\label{eq4.5}
\langle \mathsf{D}^i(x^a)\mathsf{D}^i(y^b)\rangle_{\mathcal{D}}=\delta^{ij}\frac{12B}{|x-y|^{2\Delta_{\mathsf{D}}}}\,.
\end{equation}
The quantity $B$ plays a similar role to the coefficient of the two-point function of the stress tensor for standard CFTs. Furthermore, for line defects, $B$ has a physical interpretation as the bremsstrahlung function, namely the energy emitted by a slowly moving particle \cite{Correa:2012at}. 
 
The final set of CFT data $b_{O\hat O}$ is needed to specify bulk-to-boundary OPE. This means that the configuration with the bulk operator very close to the defect is indistinguishable from an infinite sum of defect excitations. For a scalar bulk operator of dimension $\Delta$, it reads 
\begin{equation}
O(x^a,x^i)\sim b_{O\hat O}|x_i|^{\hat \Delta-\Delta}\hat O(x_a)+\cdots\,,
\end{equation}
where $\hat \Delta$ is the dimension of the operator $\hat O(x_a)$. 
In summary, we identify a set of new CFT data specified by $h_O$, $\Delta_{\hat O}$, $b_{O\hat O}$, $\hat c_{\hat O_i,\hat O_j,\hat O_k}$. However, they are not independent because crossing symmetry dictates nontrivial relationships between the different data.
The simplest observable to impose these constraints is the two-point functions of bulk operators. The systematic study of these constraints leads to the defect bootstrap program \cite{Liendo:2012hy}. 

All the above considerations do not depend on supersymmetry. In the rest of the paper, we will compute some of the introduced observables in supersymmetric theories.

\section{Topological sector and line operators}\label{sec2}

In this section we aim to identify supersymmetric configurations involving line operators and local operators. In $\mathcal{N}\ge4$ theories, a family of local protected operators was introduced in \cite{Chester:2014mea, Beem:2016cbd}. Since their correlators restricted on a line are position-independent, we will refer to them as the \emph{topological sector}. These operators are the natural candidate to play the role of local operators in our putative construction. 

To familiarize ourselves with the topological sector, we recall its construction in $\mathcal{N}\ge 4$ superconformal field theories (SCFTs). The 3d superconformal algebra is $\mathfrak{osp}(4|4)$.
\footnote{In the following, we shall work with the complexified algebras as in \cite{Beem:2013sza}. We will not discuss the physical real section since we do not directly use constraints from unitarity in Lorentzian signature.}
 Its maximal bosonic subalgebra is the direct sum 3d conformal algebra $\mathfrak{sp}(4)$ and the R-symmetry algebra $\mathfrak{su}(2)_H\oplus\mathfrak{su}(2)_C$, whose generators are denoted by ${R_{a}}^{b}$ and ${{\bar R}_{\dot a}}^{\; \; \dot b}$, respectively. The fermionic generators are given by the Poincar\`e supercharges $Q_{\alpha, a \dot a}$ and superconformal charges $S_{\alpha,a\dot a}$ \footnote{See again App \ref{appb} for more details.}.

The idea for the construction of the topological sector is to adapt the construction of the chiral ring to 3d SCFTs.
To have non-trivial correlators, we choose the relevant nilpotent supercharge $\mathcal{Q}$ to be a linear combination of the $Q_{\alpha, a \dot a}$ and $S_{\alpha,a\dot a}$ generators \cite{Beem:2013sza, Chester:2014mea}. Then, we demand the existence of a $\mathcal{Q}$-exact R-twisted translation $\hat P\sim P+R$, for some ordinary translation $P$ along a fixed direction.
Similarly to the chiral ring case, we can move $\mathcal{Q}$-closed operators in the $P$-direction by acting with $\hat P$ without changing the $\mathcal{Q}$-cohomology. 
Therefore, correlation functions of $\mathcal{Q}$-closed operators placed on the 1d submanifold generated by the translation turn out to be piecewise constant, depending at most on the order of the insertions. Unlike the chiral ring in SCFTs, these correlators do not have to vanish because the twisted translation generates R-symmetry-dependent factors, which combine with the operators to form R-symmetry singlets.

To be concrete, in 3d Euclidean flat space we take the 1d submanifold to be the line $x_1=x_2=0$ and focus on the construction of topological operators in the Higgs branch first. The maximal superconformal algebra preserved by the line is a central extension of the algebra $\mathfrak{su}(1,1|2)$, whose bosonic subalgebra includes the $\mathfrak{so}(2,1)$ conformal algebra and the preserved $\mathfrak{su}(2)$ R-symmetry algebra that we identify with $\mathfrak{su}(2)_H$. The bosonic generators are $P_{3}$, $K_{3}$, $D$, ${R_{a}}^b$, and the fermionic ones are $Q_{1,a\dot 2}$, $Q_{2,a\dot 1}$, $S_{2,a\dot 1}$, and $S_{1,a\dot 2}$. The central element is $Z=iM_{12}-R_{\dot 1\dot 2}$, where $M_{12}$ is the generator of rotations in the plane orthogonal to the line.

If we consider the two supercharges
\begin{equation}\label{eq:supercharges}
Q_1^H\equiv Q_{11\dot{2}}+\frac{1}{2r}{S^2}_{2\dot{2}}\,,\qquad Q_2^H\equiv Q_{21\dot{1}}+\frac{1}{2r}{S^1}_{2\dot{1}}\,,
\end{equation}
with $r$ being an arbitrary length parameter, from the $\mathfrak{su}(1,1|2)$ algebra it is easy to see that they are both nilpotent operators. Moreover, their anticommutator reads
\begin{equation}
\acomm*{Q_1^H}{Q_2^H}=\frac{1}{r} Z\,.
\end{equation} 
It follows that since $Z$ must vanish on the $Q^H_{1,2}$ cohomology classes, operators belonging to these classes are inserted along the fixed point locus of $M_{12}$ and have zero $R_{\dot 1\dot 2}$ charge. 

We now perform a topological twist by combining the $\{P_3, K_3, D\}$ generators of the $\mathfrak{so}(2,1)$ conformal algebra along the line and the $\mathfrak{su}(2)_H$ R-symmetry generators. 
They are given by
\begin{align}
\hat{P}=P_3-\frac{1}{2r}{R_{2}}^{1}\,,\qquad \hat{K}=K_3-2r{R_{1}}^{2}\,,\qquad  \hat{D}=D+{R_1}^{1}\,.
\end{align}
One can think of them as the generators of a new one-dimensional conformal algebra along the $x_3$-line.
It turns out that the twisted generators are all $Q_{1,2}^H$-exact. Therefore, we can first construct operators localized at the origin from $Q_{1,2}^H$-closed, gauge invariant operators of the 3d theory. 
Explicitly, the cohomology of the two supercharges contain local operators $O_{a_1,\,\dots,\,a_n}$ with the following properties \cite{Chester:2014mea}: they are Lorentz scalars, transform in the $(\mathbf{n+1},\,\mathbf{1})$ of  $\mathfrak{su}(2)_H\oplus\mathfrak{su}(2)_C$, and have conformal dimension $\Delta=-n/2$. We then move them along the line by applying the twisted translation generator $\hat{P}$. The corresponding twisted translated operator at position $\vec{x}=(0,\,0,\,s)$ is given by
\begin{equation}
O(s)=e^{i s \hat{P}}O_{1,\dots,1}(0)e^{-i s \hat{P}}=O_{a_1,\dots,a_n}(\vec x)\big|_{\vec x=(0,0,s)} u^{a_1}\dots u^{a_n}\,, \qquad u^a=\left(1,\frac{is}{2r}\right)\,.
\end{equation}
These operators are still $Q_{1,2}^H$-closed and form the Higgs topological sector of the ${\cal N} \geq 4$ SCFT on the line.
An analogous construction can be carried on for Coulomb branch operators by exchanging dotted and undotted indices. This amounts to exchange the role of $\mathfrak{su}(2)_C$ and $\mathfrak{su}(2)_H$.

\vskip 10pt

One of the crucial insights of \cite{Dedushenko:2016jxl} is that the topological sector can also be defined away from the CFT point. In the end, what is needed is a couple of supercharges $Q$, $\tilde{Q}$ such that $\acomm*{Q}{\tilde{Q}}$ yields the twisted translation. We also require $Q^2$ to close on a different twisted translation, which leaves invariant the locus generated by $\acomm*{Q}{\tilde{Q}}$. Thus we can define a sensible $Q$-cohomology. As in the conformal case, operators in the $Q$-cohomology define the topological sector. 
For UV gauge theories on $S^3$, one can define topological sectors on any great circle of $S^3$.
If the UV gauge theories are good in the sense of \cite{Gaiotto:2008ak}, the UV topological sectors will contain physical information on the IR SCFT.

In the following, we generalize this construction to have an additional BPS loop operator on the great circle fixed by $Q^2$, and with a topological sector on the linked great circle. 
For SCFTs, we can map the setup to flat space, where the defect extends along a straight line direction, and the topological sector lives on the circle of radius $r$ in the orthogonal plane. 
We first introduce our construction for the Wilson and vortex loops in UV gauge theories defined in Sec.~\ref{sec1.1}. 
These are the simplest examples in which our construction is available. Then, we apply it to the fermionic Wilson line in ABJM. From this example, we argue the existence of a topological sector compatible with any $\frac{1}{2}$--BPS line operators in $\mathcal{N}=4$ SCFTs.

\subsection{Decorating the Wilson loop}

The first example we investigate is the Wilson loop in UV gauge theories on $S^3$, defined in Eq.~\eqref{uvwl}. The explicit preserved $\mathfrak{su}(1|1)\oplus \mathfrak{su}(1|1)$ algebra is
\begin{align}
\acomm*{Q_2^{(\ell+)}}{Q_1^{(\ell-)}}&=\frac{i}{r}\left(J_{12}^{(\ell)}+\frac{R_C-R_H}{2} \right)\,,\\
\acomm*{Q_2^{(r+)}}{Q_1^{(r-)}}&=\frac{i}{r}\left(J_{12}^{(r)}+\frac{R_H+R_C}{2} \right)\,,
\end{align}
where $J_{12}^{(\ell)}$ and $J_{12}^{(r)}$ are generators of isometries on $S^3$ (see App.~\ref{appb}) and their action on operators is given by
\begin{equation}
P_\tau=i\partial_\tau=J_{12}^{(\ell)}+J_{12}^{(r)}\,, \qquad  P_\varphi=i\partial_\varphi=J_{12}^{(\ell)}-J_{12}^{(r)}\,,
\end{equation}
where $P_\tau$ and $P_\varphi$ generates translations along the $\tau$ and $\varphi$ circle, respectively.
$R_H$ and $R_C$ are R-symmetry transformations generated by $R_H=\frac{1}{2}{h_{a}}^{b}{R_{b}}^{a}$ and $R_C=\frac{1}{2}{\bar h_{\dot a}}^{\;\;\dot b}{\bar R_{\dot b}}^{\;\;\dot a}$ and acts on operators as 
\begin{equation}
\comm{R_H}{O_a}=-\frac{1}{2}{h_{a}}^bO_b\,,\qquad
\comm{R_C}{O_{\dot a}}=-\frac{1}{2}{\bar h_{a}}^{\dot b}O_{\dot b}\,.
\end{equation}

We are looking for a one-parameter family of supercharges squaring on a linear combination of $P_\varphi$ and an R-symmetry transformation which leaves the Wilson loop invariant, that is $R_H$. We choose our supercharge to be a linear combination of the nilpotent supercharges
\begin{equation}\label{wlqcoh}
Q^{\textup{W}}_1=Q_2^{(\ell+)}+iQ_1^{(r-)}\,, \qquad Q^{\textup{W}}_2=Q_1^{(\ell-)}+iQ_2^{(r-)}\,.
\end{equation}
Then, the ``cohomological'' supercharge is $Q_\beta^{\textup{W}}=Q^{\textup{W}}_1+\beta \,Q^{\textup{W}}_2$ and satisfies 
\begin{equation}
(Q_\beta^{\textup{W}})^2\propto P_\varphi-R_H\,.
\end{equation} 
We are looking for a family of topological operators annihilated by $Q_\beta^{\textup{W}}$. As in the case without defect, we interpret $P_\varphi-R_H$ as the central charge $Z$. Consequently, the operators in the cohomology of $Q_\beta^{\textup{W}}$ must be annihilated by $P_\varphi-R_H$. Then, they must be Coulomb branch operators and placed on $S^1_\tau$.

It is not hard to construct the corresponding twisted translation $\hat P_\tau$. Because of the following anticommutators
\begin{align}
\acomm*{Q^{\textup{W}}_1}{Q^{(\ell-)}_1 - i Q^{(r+)}_2}=\acomm*{Q^{\textup{W}}_2}{{Q^{(\ell+)}_2 - i Q^{(r-)}_1}}=\frac{i}{r}\left(P_\tau+R_C\right)\,,
\end{align}
we choose it to be $\hat P_\tau=P_\tau+R_C$.
We have thus recovered the necessary minimal structure for the topological sector. Unlike in the case of SCFT, we cannot rely on representation theory to classify the cohomology of $Q_\beta^{\textup{W}}$. Therefore, we need to check explicitly whether an operator is in the cohomology or not. If we restrict ourselves to operators constructed from fundamental fields, we find that \footnote{The cohomology contain also interesting BPS monopole operators like in \cite{Dedushenko:2017avn, Dedushenko:2018icp}, but we do not address that problem in this paper.}
\begin{equation}
\Phi(0)=\frac{1}{2}\Phi_{\dot 1\dot 1}(0)- \Phi_{\dot 1\dot 2}(0)+\frac{1}{2}\Phi_{\dot 2\dot 2}(0)
\end{equation}
is annihilated by $Q^\textup{W}_i$ at $\theta=0$, $\tau=0$. We use the twisted translation $\hat{P}_\tau=P_\tau+R_C$ to move $\Phi$ along the $\tau$-circle. The twisted translated operator is annihilated by $\mathcal{Q}_\beta^{\textup{W}}$ and it reads 
\begin{equation}\label{coulop}
\Phi(\tau)=v^{\dot a}(\tau)v^{\dot b}(\tau)\Phi_{\dot a\dot b}(\tau)\,, \qquad v^{\dot a}(\tau)=\frac{1}{\sqrt{2}}\left(e^{\frac{i}{2}\tau}, -e^{-\frac{i}{2} \tau}\right) \,.
\end{equation}
The gauge invariant polynomials of $\Phi(\tau)$ are the simplest topological operators compatible with the BPS Wilson loop.

\subsection{Decorating the vortex loop}

In this section, we study configurations involving vortex loops defined in Eq.~\eqref{vl} and topological operators. Based on mirror symmetry \cite{Intriligator:1996ex}, which exchanges Wilson and vortex loops and the Higgs branch protected sector with the Coulomb branch one \cite{Dedushenko:2017avn}, we expect the existence of a supersymmetric setup in which the vortex loop is on $S^1_\varphi$ and the Higgs branch topological sector is on $S^1_\tau$ \footnote{
If we consider theories described by a Hanany-Witten construction within type IIB string theory \cite{Hanany:1996ie}, both local and extended operators can be realized through specific brane configurations \cite{Assel:2015oxa, Assel:2017hck}. 
 Consequently, it is reasonable to expect the existence of a brane construction engeneering our combined system. Considering that mirror symmetry acts locally on these branes \cite{Gaiotto:2008ak}, we expect the two setups to exhibit mirror symmetry. However, explicit checks of this conjecture lie beyond the scope of this paper.}. 
We briefly check that the intuition is indeed correct.

The preserved $\mathfrak{su}(1|1)\oplus\mathfrak{su}(1|1)$ algebra is
\begin{align}
\acomm*{Q_2^{(\ell+)}}{Q_1^{(\ell-)}}&=\frac{i}{r}\left(J_{12}^{(\ell)}+\frac{R_C-R_H}{2} \right)\,,\\
\acomm*{Q_1^{(r+)}}{Q_2^{(r-)}}&=\frac{i}{r}\left(J_{12}^{(r)}-\frac{R_H+R_C}{2} \right)\,.
\end{align}
We follow the same logic as before, but we exchange the role of $R_C$ and $R_H$. In this way, we find the nilpotent supercharges are
\begin{equation}
\mathcal{Q}^{\textup{V}}_1= Q_2^{(\ell+)}+ iQ_1^{(r+)}\,, \qquad
\mathcal{Q}^{\textup{V}}_2=Q_1^{(\ell-)}+ iQ_2^{(r-)}\,.
\end{equation}
Their anticommutator plays again the role of the central charge
\begin{equation}
\acomm*{\mathcal{Q}^{\textup{V}}_1}{\mathcal{Q}^{\textup{V}}_2}=\frac{i}{r}\left(P_\varphi+R_C \right)\,.
\end{equation}
Then, a putative topological sector can only live on $S^1_\tau$ and contains Higgs branch operators. Thus, the corresponding twisted translation is easily derived from
\begin{align}
\acomm*{\mathcal{Q}^{\textup{V}}_1}{Q_1^{(\ell-)}- iQ_2^{(r-)}}=
\acomm*{\mathcal{Q}^{\textup{V}}_2}{Q_2^{(\ell+)} - iQ_-^{(r+)}}&=\frac{4i}{r}\left(P_\tau-R_H\right)\,,
\end{align}
and it reads $\hat{P}_\tau=P_\tau-R_H$.
The natural candidates for a non-trivial cohomology are the scalar fields in the hypermultiplet.
By direct inspection we find that both $q^1$ and $\tilde{q}^1$ are annihilated by any linear combination of $\mathcal{Q}^{\textup{V}}_1$ and $\mathcal{Q}^{\textup{V}}_2$. Then, we build the corresponding twisted fields by acting with the twisted translation $\hat P_\tau=P_\tau-R_H$ 
\begin{equation}\label{higgsop}
\mathfrak{q}(\tau)\equiv u^a(\tau) q_a(\tau), \quad \tilde{\mathfrak{q}}(\tau)\equiv u^a(\tau) \tilde q_a(\tau)\,, \quad u^a=\left(-\sin\frac{\tau}{2},-\cos\frac{\tau}{2}\right)\,.
\end{equation}
Thus, the Higgs branch topological sector is formed by the gauge invariant polynomials of the twisted fields $\mathfrak{q}$, $\tilde{\mathfrak{q}}$.

\subsection{The superconformal case}\label{sec2.3}

We now concentrate on superconformal lines in SCFTs with at least $\mathcal{N}=4$ supersymmetry. While the construction in UV gauge theories relies on the existence of a Lagrangian description, our goal is to extend it to line operators regardless of their precise definition or that of the SCFT. However, for simplicity and with an eye to the applications in the following sections, we begin with a detailed discussion of the fermionic Wilson line in ABJM \cite{Drukker:2009hy}. Building on this example, we will argue a more general conclusion for $\mathcal{N}=4$ SCFTs.

To begin with, we briefly describe the $\mathfrak{su}(1,1|3)\oplus \mathfrak{u}(1)_b$ superalgebra preserved by the fermionic Wilson loops \cite{Bianchi:2017ozk}. 
The Wilson loop preserves the following  supercharges 
\begin{equation}
\bar Q_{1,34}\,, \; \bar Q_{1,32}\,, \; \bar Q_{1,24}\,, \; \bar Q_{2,14}\, \; \bar Q_{2,12}\,,\; \bar Q_{2,13}\,,\; \bar S_{1,34}\,\;\bar S_{1,32}\,,\;\bar S_{1,24}\,\;\bar S_{2,14}\,,\; \bar S_{2,12}\,,\;\bar S_{2,13}\,,
\end{equation}
where $\bar Q_{\alpha,IJ}$ and $\bar S_{\alpha,IJ}$ are Poincar\'e and superconformal superchages respectively, and $I,J=1,\dots,4$ are the antisymmetric $\mathfrak{su}(4)$ indices.
They close on the 1d conformal group spanned by $P_3$, $K_3$, and $D$, and on the $\mathfrak{su}(3)$ R-symmetry generators ${L_a}^b$, whose precise description is in the appendix. The additional $\mathfrak{u}(1)_b$ factor commutes with all these generators, and we can safely ignore it for the rest of the paper. 

We want to make contact with the algebra of the Wilson loop in UV gauge theories described starting from Eq.~\eqref{wlqcoh}.  As a first step, we break down the full $\mathcal{N}=6$ superconformal algebra to the $\mathcal{N}=4$ algebra. We decide to identify the $\mathfrak{su}(2)_H$ factor of the residual $\mathcal{N}=4$ R-symmetry algebra with an $\mathfrak{su}(2)$ factor of the algebra preserved by the fermionic Wilson loop. We denote these generators as ${R_a}^b$. Then, we construct the Poincar\'e subalgebra $\mathfrak{su}(2|1)_\ell\oplus\mathfrak{su}(2|1)_r$ in the usual way. The details are rather boring and technical, and the interested reader can find them in the App.~\ref{appb}. The upshot is that we can embed the supercharges preserved by the Wilson loop in the relevant Poincar\'e superalgebra. 
That is, if we take the supercharges $Q^\textup{W}_1$ and $Q^\textup{W}_2$ defined in \eqref{wlqcoh}, and we use the explicit embedding derived in App.~\ref{appb}, we find the two cohomological supercharges
\begin{align}
Q_1&=\frac{1}{2}\left[i \bar Q_{1,32}-\bar Q_{1,34}+\bar Q_{2,12}+i \bar Q_{2,14}+\frac{1}{2r}\left(i \bar S_{1,32}-\bar S_{1,34}-\bar S_{2,12}-i \bar S_{2,14}\right)\right]\,,\\
Q_2&=\frac{1}{2}\left[\bar Q_{1,32}-i \bar Q_{1,34}+i\bar Q_{2,12}+\bar Q_{2,14}+\frac{1}{2r}\left(i \bar S_{1,34}+i \bar S_{2,12}+\bar S_{2,14}-\bar S_{1,32}\right)\right]\,.
\end{align}
They annihilate the Wilson loop, and their anticommutator closes on $P_\varphi-R_H$, where we are implicitly compactifying the theory on $S^3$. Moreover, we can define the twisted translations from the anticommutators $\acomm*{Q_1}{\tilde Q_1}$ and $\acomm*{Q_2}{\tilde Q_2}$, where
\begin{align}
\tilde Q_1&=\frac{1}{2}\left[ \bar Q_{1,32}-i \bar Q_{1,34}-i \bar Q_{2,12}-\bar Q_{2,12}+\frac{1}{2r}\left(i \bar S_{1,34}-\bar S_{1,32}-i \bar S_{2,14}-\bar S_{2,14}\right)\right]\,,\\
\tilde Q_2&=\frac{1}{2}\left[ -i\bar Q_{1,32}+\bar Q_{1,34}+\bar Q_{2,12}+i \bar Q_{2,12}+\frac{1}{2r}\left(\bar S_{1,34}-i \bar S_{1,32}-\bar S_{2,14}-i \bar S_{2,14}\right) \right]\,.
\end{align}
Then, the twisted translation is again $P_\tau-R_C$, being $R_C$ a combination of the original $\mathfrak{su}(4)$ R-symmetry generators.

The final step is to examine the cohomology. In the conformal case, this problem has been already addressed on the line in full generality in \cite{Chester:2014mea} for $\mathcal{N}=4$ SCFTs. Then, up to a conformal transformation that does not affect the cohomology, we do not need to repeat their analysis. Here we limit ourselves to writing down the operator, which will be studied in detail in the following sections. A natural candidate is the gauge invariant polynomials of the scalar fields $C_I$, $\bar C^I$. 
Indeed, studying their supersymmetric variation \eqref{susytransf} w.r.t. to any linear combination of $Q^\textup{W}_1$ and $Q^\textup{W}_2$, we find that the combinations $C_1+C_3$ and $\bar C^1-\bar C^3$ vanishes for $\tau=0$. Then, with the now familiar procedure, we identify the building blocks for gauge invariant topological operators
\begin{equation}
\mathcal{C}(\tau)=U^I(\tau)C_I(\tau)\,,\qquad
\bar{\mathcal{C}}(\tau)=\bar V_I(\tau)\bar C^I(\tau)\,,
\end{equation}
with
\begin{equation}
U^I(\tau)=\frac{1}{\sqrt{2}}\left(e^{-\frac{i}{2}\tau},\,0\,,\,e^{\frac{i}{2}\tau},\,0\,\right)^I\,,\qquad
\bar V_I(\tau)=\frac{1}{\sqrt{2}}\left(e^{\frac{i}{2}\tau},\,0\,,\,-e^{-\frac{i}{2}\tau},\,0\,\right)_I\,.
\end{equation}
For instance, the simplest operator we can study is
\begin{equation}\label{myop}
\mathcal{O}(\tau)=\Tr(\bar{\mathcal{C}}(\tau)\mathcal{C}(\tau))\,,
\end{equation}
which is also part of the stress tensor multiplet \footnote{The fact that the stress tensor multiplet admits a protected topological operator is a special feature of $\mathcal{N}=6$ SCFTs. In a generic $\mathcal{N}=4$ SCFT the dimension one topological operator is part of a conserved current multiplet. }. 

As a check of the correctness of our results, we can see that it agrees with 3d IR dualities. ABJM theory with gauge group $U(N)_1\times U(N)_{-1}$ and Chern-Simons level $k=1$ is dual to a UV Yang-Mills theory coupled to one fundamental and one adjoint hyper \cite{Kapustin:2010xq}. 
From an $\mathcal{N}=4$ SUSY perspective, the operator $\mathcal{O}(\tau)$ is the bilinear built out of a twisted-hypermultiplet, that is a hypermultiplet with the $\mathfrak{su}(2)_C$ and $\mathfrak{su}(2)_H$ exchanged\footnote{More precisely, we can think of $C_1$, $C_3$ as the $\mathfrak{su}(2)_C$-doublet of a twisted-hypermultiplet transforming in a representation $R$ of the gauge group. That is, we embed $C_1$, $C_3$ into a scalar field $q_{\dot a}$. Similarly, $\bar C^1$, $\bar C^3$ are embedded into the doublet $\tilde q^{\dot a}$, transofrming in the complex conjugate representation of $q_{\dot a}$.}.
After the duality transformation, this operator is mapped to the Coulomb branch operator built out of $\Phi_{\dot a\dot b}$ introduced in \eqref{coulop} \cite{Hayashi:2022ldo}, which is also part of the cohomology in the dual theory. In other words, at the level of local operators, the cohomologies are mapped consistently under the duality. At the level of loop operators, even without the insertion of additional local operators, the duality is not fully understood yet, and we refer to \cite{Griguolo:2021rke} for recent progress. 

Now we are ready to generalize our construction to generic $\mathcal{N}=4$ SCFTs.
Our argument is based on the superconformal algebra rather than its specific realization in a given theory, and it extends to all $\frac{1}{2}$--BPS line operators in $\mathcal{N}=6$ SCFTs. Moreover, we have not used the full $\mathcal{N}=6$, but only its $\mathcal{N}=4$ subalgebra. Similarly, for the Wilson line, we have not exploited the full $\mathfrak{su}(3)$ R-symmetry algebra, but only an $\mathfrak{su}(2)$ factor.
We conclude that our construction can be easily extended to any BPS line operators in $\mathcal{N}=4$ SCFT preserving an $\mathfrak{su}(2)$ factor. Looking at the classification of the superconformal lines in 3d SCFTs of \cite{Agmon:2020pde}, we conclude that there is a topological sector compatible with any $\frac{1}{2}$--BPS line operators in $\mathcal{N}=4$ SCFTs. This is the first main result of the paper. In the following, we discuss how to compute defect CFT data from our setup.


\section{Exact formula for stress tensor correlators in ABJM}\label{sec3}

In this section, we present a formula for extracting CFT defect data for BPS lines. Our method is based on a powerful Ward identity that relates topological operators belonging to conserved current multiplets to mass or FI -deformations of the $S^3$ partition function. Since the latter is often amenable to supersymmetric localization, many exact results are accessible.
Elaborating on this idea further, we will argue that the vev of the defect in the properly deformed background computes the defect correlation functions of the protected operators. Finally, we will apply our formula to the maximally supersymmetric Wilson line in ABJM.

\subsection{The cohomological Ward identity}

We review the relation between the topological sectors and supersymmetric deformations of the $S^3$ partition functions. Let us begin with mass deformations. 
They can be built whenever a theory has a flavor symmetry by using the corresponding current multiplet $\Sigma^A =(J^A_{ab},\, \chi^A_{a\dot{a}}, \, j^A_\mu, \, K^A_{\dot{a}\dot{b}})$.\footnote{Here $A$ is an index which runs from 1 to the rank of the flavor symmetry Lie algebra.} The dimension-one scalars $J^A_{ab}$ are in the $(\mathbf{3},\mathbf{1})$ of the R-symmetry group, $\chi^A_{a\dot{a}}$ are the fermion partners of dimension $3/2$ in the $(\mathbf{2},\mathbf{2})$ of the R-symmetry group, $j_\mu^A$ are the flavor conserved currents, and $ K^A_{\dot{a}\dot{b}}$ are dimension-two scalars in the $(\mathbf{1},\mathbf{3})$ of the R-symmetry group. A current multiplet can always be coupled to a background vector field. If we take the supersymmetric background in which
$\Phi_{\dot a\dot b}=m\bar h_{\dot a\dot b}$, $D_{ab}=h_{ab}m/r$, and all the other fields vanishing, the corresponding action is a real mass deformation. 
This amounts to modifying the action by the following terms
\begin{equation}\label{massdef}
S_{\textup{mass}}=m_A \int_{S^3} d^3x\sqrt{g}\left(-ih^{ab}J^A_{ab}+\bar h^{\dot a\dot{b}}K^A_{\dot{a}\dot{b}}\right) +O\left(m^2\right) \,.  
\end{equation}
The terms of order $m^2$ are needed to preserve supersymmetry, but its explicit expression will not be important for us.
We also observe that $J^A_{ab}$ has the quantum numbers of a Higgs branch operator, and therefore it can be made topological. The corresponding twisted operator is $J^A(\tau) = J^A_{ab}u^au^b$. Thus, we may expect a relation between mass terms and dimension one topological operators. Building on this intuition, it turns out that the mass deformation $S_{\textup{mass}}[m]$ is almost a total supersymmetric variation w.r.t. to the cohomological supercharge preserving the topological sector. The only non-exact part is a boundary term, namely the ``topologized'' version of the superconformal primary of the current multiplet integrated over the great circle supporting the topological sector \cite{Guerrini:2021zuk, Bomans:2021ldw}.  In formulae we get
\begin{equation}\label{Higgs:fundeq}
\pdv{m}\left(S_{\textup{mass}}[m]-4\pi r^2 m\oint_{S^1_{\varphi}}    J(\varphi)   \right)= \acomm{Q_\beta^H}{\dots}\,,
\end{equation}
where the explicit expression in the right entry of the anticommutator is not necessary and can be found in \cite{Guerrini:2021zuk}. 
It is not hard to use the Ward identity to prove that the mass deformed partition function is the generating functional of integrated topological operators, that is
\begin{equation}\label{conjecture}
\hspace{-0.3cm} \Big\langle \left(\int d\varphi_1\, J^{A_1}(\varphi_1)\right)^{k_1}\dots \left(\int d\varphi_n \,J^{A_n}(\varphi_n)\right)^{k_n}\Big\rangle=\left( -\frac{1}{4\pi r^2} \right)^{\! n} 
\, \frac{1}{Z}\frac{\partial^k Z[m_{A_1},\dots ,m_{A_n}]}{\partial m^{k_1}_{A_1}\dots \partial m^{k_n}_{A_n}}\bigg|_{m_{A_1} \!, m_{A_2} \!, \cdots = 0}\,,
\end{equation}
where $k=k_1+\dots+k_n$ with $k_1, \dots, k_n\in\mathbb{N}$. This version of the Ward identity is very useful for calculating correlation functions and extracting CFT data. For instance, the r.h.s. can be evaluated exactly with localization, and related to the non-trivial CFT data stored in the l.h.s.. A remarkable physical application of this method is the evaluation of the coefficients of the string theory and M-theory action beyond the SUGRA limit \cite{Agmon:2017xes, Binder:2019mpb, Binder:2020ckj}.

The localization formula for topological correlations function in UV gauge theories of \cite{Dedushenko:2016jxl} gives a highly non-trivial consistency check. According to \cite{Dedushenko:2016jxl}, topological correlators are captured by a quadratic quantum mechanics. The resulting 1d action $S_\sigma[Q\,,\tilde{Q}]$ is coupled via a mass term to the standard matrix model arising from the usual localization scheme for the vector multiplet. 
The quantum mechanics' fundamental fields $Q,\,\tilde{Q}$ of the effective theory are identified with the twisted translated operators built out of the hypermultiplet (see Eq.~\eqref{higgsop}). Now we turn on the mass deformation of Eq.~\eqref{massdef}. The localized expression gets modified by adding a 1d mass term for the 1d action
\begin{equation}\label{HBO:loc}
S_\sigma[Q\,,\tilde{Q}]\to S_\sigma[Q\,,\tilde{Q}]- 4\pi r^2 m^A \int_{-\pi}^\pi d\varphi \,\tilde{Q}(\varphi)T^A Q(\varphi)\,,
\end{equation}
where $T^A$ are the generators of the flavor symmetry in the proper representation. The operators $J^A(\varphi)\equiv - \tilde{Q}(\varphi)T^A Q(\varphi)$ are the twisted operators $J^A(\varphi) $ is nothing but the lowest component of the current multiplet coupled to the background vector field. Then, derivatives w.r.t. to $m$ reproduces \eqref{conjecture}.

The Ward identity does not require any localization but only the existence of a flavor current.
Therefore, the range of validity of \eqref{conjecture} is extended to any $\mathcal{N}=4$ SCFTs, regardless of their specific realization. These include even theories where the localization argument has not been developed yet, like Chern-Simons matter theories. This observation is crucial to apply the formula to ABJM theory.

The new step is to apply this identity in the presence of a defect $\mathcal{D}$ preserving $Q$. Until the insertion is $Q$-closed, the cohomological argument goes through. Thus we can still use Eq.~\eqref{Higgs:fundeq} to relate correlation functions of topological operators with the defect $\mathcal{D}$ to mass derivatives of the vev of the defect with a mass-deformed backgrounds. 
Extending the explicit formula is straightforward 
\begin{equation}\label{defect:formula}
\hspace{-0.3cm} \Big\langle \left(\int d\varphi_1\, J^{A_1}(\varphi_1)\right)^{k_1}\dots \left(\int d\varphi_n \,J^{A_n}(\varphi_n\right)^{k_n}\Big\rangle_\mathcal{D}=\left( -\frac{1}{4\pi r^2} \right)^{\! n} 
\, \frac{1}{\mathcal{D}}\frac{\partial^n \langle \mathcal{D}\rangle[m_1,\dots ,m_n]}{\partial m^{k_1}_{A_1}\dots \partial m^{k_n}_{A_n}}\bigg|_{m^{A_1} \!, m^{A_2} \!, \cdots = 0}\,,
\end{equation}
where $\langle \mathcal{D}\rangle$ is the vev of the defect and $\langle \rangle_{\mathcal{D}}$ indicates the normalized correlation function in the presence of the defect. For instance, we can use \eqref{defect:formula} to compute correlation functions of Higgs branch operators with the vortex loop introduced in Eq.~\eqref{vl}. However, as in the case without the defect, our formula is independent of the explicit realization of the theory and the defect.

In an analogous way, we can extend the formula to Coulomb branch operators, even in the presence of a BPS defect. Here the relevant deformation is the Fayet-Iliopoulos (FI) term.
To write it down, we think of $\Phi_{\dot a \dot b}$ as the bottom component of the current multiplet of the topological current $j_\mu\propto\epsilon_{\mu\nu\rho}F^{\nu\rho}$. The multiplet also includes the fermions $\lambda_{a\dot a}$ and the auxiliary fields $D_{ab}$. For each $U(1)$ factor of the gauge group, there is a multiplet of this type. One can couple this multiplet to an abelian background twisted vector multiplet $\tilde{\mathcal{V}}_{\rm back}$. The relevant coupling is simply an $\mathcal{N}=4$ mixed Chern-Simons term \footnote{For details see the App.~B of \cite{Guerrini:2021zuk}.}. In the rigid limit, we get the FI term on $S^3$
\begin{equation}\label{FI}
S_{FI}=i\zeta\int_{S^{3}} d^3 x\sqrt{g}\left({h_a}^b{D_b}^a-\frac{1}{r}\bar{h}\indices{^{\dot{a}}_{\dot{b}}} \Phi\indices{^{\dot{b}}_{\dot{a}}}\right)\,.
\end{equation}
As discussed in \cite{Guerrini:2021zuk}, we can write a similar Ward identity for such a deformation following the same logic of real masses 
\begin{equation}\label{Ward:coul}
S_{FI}=-4\pi ir^2\int_{S^1_\varphi}d\varphi\,\Phi_{\dot{a}\dot{b}}v^{\dot{a}}v^{\dot{b}}+\delta_\xi\left[ \int_{S^3} d^3x \sqrt{g}\,\tilde\eta^{a\dot a}\lambda_{a\dot a}\right]\,.
\end{equation}
Then, derivatives w.r.t. to the FI parameters gives 1d integrated Coulomb branch operators. Even in this case, the formula is compatible with localization for Coulomb branch operators. Again, any defect preserving the cohomological supercharge can be inserted without additional complications. In this way, we get another identity for defect correlation functions
\begin{equation}\label{cbo:equiv}
\langle \int_{S^1_{\varphi_1}}d\varphi_1\,\Phi(\varphi_1)\cdots\int_{S^1_{\varphi_n}}d\varphi_n\,\Phi(\varphi_n)\rangle_\mathcal{D}=\left(\frac{i}{4\pi r^2}\right)^n\frac{\partial^n}{\partial \zeta^n}\langle \mathcal{D}\rangle[\zeta]\,.
\end{equation}
In UV gauge theories, Eq.~\eqref{cbo:equiv} allows us to compute the corresponding defect topological correlators in the presence of the Wilson loop of Eq.~\eqref{uvwl}. However, we stress again that these types of formulas are independent of the specific description of the theory and the defect, and hold again for any $\frac{1}{2}$--BPS conformal line operators in any $\mathcal{N}=4$ SCFTs.

In the rest of the section, we apply our formula to the other explicit example we discussed, namely $\frac{1}{2}$--BPS Wilson loops ABJM.

\subsection{Defect correlation functions in ABJM}\label{sec3.2}

In Sec.~\ref{sec2.3} we showed the existence of a supersymmetric configuration involving the fermionic Wilson line and the topological sector in ABJM. Even if a localization scheme in Chern-Simons matter theory is missing, we can still apply the Ward identity. 
All that is left is to identify the correct deformation coupled to the current multiplet having as the superconformal primary the operator
\begin{equation}\label{opabjm}
{O_I}^J=\Tr(C_I\bar C^J)-\frac{1}{4}\delta_I^J\Tr(C_K\bar C^K)\,.
\end{equation}
The case of ABJM is particularly interesting as the bilinear operator is part of the stress tensor multiplet. Then, Ward identities can relate correlation functions of twisted operators $\mathcal{O}(\tau)$ defined in Eq.~\eqref{myop} to stress tensor correlators. For instance, as we will discuss in detail later, the one-point function of $\mathcal{O}(\tau)$ will be related to the bremsstrahlung function, which is a universal defect CFT data.

As discussed at the end of Sec.~\ref{sec2.3}, looking at ABJM as a theory preserving the $\mathcal{N}=4$ supersymmetry algebra described in App.~\ref{appb}, $C_1$, $C_3$ form the $\mathfrak{su}(2)_C$ doublet of a twisted-hypermultiplet. 
Combining with its complex conjugate doublet $\bar C^1$, $\bar C^3$, we construct the bilinear operator transforming in the adjoint of $\mathfrak{su}(2)_C$, that is nothing but the operator defined in \eqref{opabjm} with indices $I$, $J$ restricted to be 1 and 3.
From the $\mathcal{N}=4$ perspective, that operator is the primary of a flavor current multiplet. Then, it can be coupled to a rigid background vector, and the resulting mass term looks like \eqref{massdef}, with dotted and undotted indices exchanged \footnote{This exchange reflects that we choose the duality frame for the $\mathcal{N}=4$ subalgebra of ABJM such that the topological sector is built out of the twisted hypermultiplet. Then, we interpret the corresponding mass term as a Coulomb branch deformation for that subalgebra, even if \eqref{massdef} is usually a Higgs branch deformation. }.

Every deformation compatible with the topological sector satisfies a cohomological Ward identity which relates derivatives w.r.t. to the deformation parameter of the action to 1d integrated topological operators.
Remarkably, this property remains invariant regardless of the specific Lagrangian realizations of the model. Then, we can readily extend it to ABJM with the mass deformation for the twisted hyper.
In our conventions, the mass term generates a Coulomb branch deformation, and hence we implement the formula \eqref{cbo:equiv}.
Consequently, the expression for integrated correlators becomes 
\begin{equation}\label{defect:formula}
\hspace{-0.3cm} \Big\langle \int d\tau_1\, \mathcal{O}(\tau_1)\dots \int d\tau_n \,\mathcal{O}(\tau_n)\Big\rangle_W=\left( \frac{i}{4\pi r^2} \right)^{\! n} 
\, \frac{1}{W}\frac{\partial^n \langle W\rangle[m]}{\partial m^n}\bigg|_{m = 0}\,,
\end{equation}
where $ \langle W\rangle[m]$ denotes the vev of the Wilson loop in the mass deformed ABJM theory on $S^3$.

The crucial point is that the l.h.s. can be computed exactly with localization.\footnote{ From the $\mathcal{N}=2$ perspective \cite{Benna:2008zy}, we are decomposing $C_I$ in terms of the $\mathcal{N}=2$ chirals $W_i$, $Z_i$ as follows: $C_I=(W_1, W_2, Z^\dagger_1, Z^\dagger_2)$. Then, if we give mass $m$ to $W_1$ and $Z^\dagger_1$ we get the desired mass deformation for $C_1$ and $C_3$.} The localized partition function reads \cite{Kapustin:2009kz, Jafferis:2010un, Hama:2010av}
\begin{align}\label{eq:matmodabjm}
Z_{\textup{ABJM}}=\frac{1}{N_1!N_2!}\int d\lambda_id\mu_j e^{\mbox{\scriptsize$ik\pi\left(\displaystyle\sum_i^{N_1}\lambda_i^2-\sum_j^{N_2}\mu_j^2\right)$}}\mbox{\footnotesize$\frac{\displaystyle\prod_{i\neq j}^{N_1}2\sinh\pi(\lambda_i-\lambda_j)\prod_{i\neq j}^{N_2}2\sinh\pi(\mu_i-\mu_j)}{\displaystyle\prod_{i,\, j}\left[2\cosh\pi(\lambda_i-\mu_j)\right]\left[2\cosh(\pi(\lambda_i-\mu_j+mr))\right]}$}\,.
\end{align}
The vev of supersymmetric operators is computed by insertions of operators in this matrix model. For the $\frac{1}{2}$--BPS Wilson loop in the fundamental representation the insertion is \cite{Drukker:2009hy}
\begin{equation}
\langle W\rangle= \langle\sum_{i=1}^{N_1} e^{2\pi\lambda_i}+\sum_{j=1}^{N_2}e^{2\pi\mu_j} \rangle\,.
\end{equation}

\paragraph{The weak coupling computation}

We find the matrix model insertion by computing derivatives in the matrix model.
For the one-point function, we obtain
\begin{equation}
\pdv{m}\langle W[m] \rangle=\langle-\sum_{i,j}\pi r\tanh\pi(\lambda_i-\mu_j)\rangle_W\,,
\end{equation}
where $\langle \rangle_W$ indicates the matrix model average with the WL insertion.

We exploit our relation at weak coupling, namely in the limit $k\gg1$. 
To expand the matrix model, we rescale the variable by a factor $\sqrt{k}$
\begin{equation}
x_i=\pi\sqrt{k}\lambda_i\,,\qquad  y_j=\pi\sqrt{k}\lambda_j\,.
\end{equation}
We also reconstruct the $U(N_1)\times U(N_2)$ Haar measure.\footnote{We recall the Haar measure for a single $U(N)$ factor 
\begin{equation}
\int dU f(U)=\int du_i \prod_{i<j} (u_i-u_j)^2 f(U)\,,
\end{equation}
where $u_i$ are the eigenvalues of $U$.}
Then, the undeformed matrix model reads as
\begin{align}
Z=\frac{1}{N_1!N_2!}\int& dXdY \,e^{i\left(\sum_{i=1}^{N_1}x_i^2-\sum_{j=1}^{N_2}y_j^2\right)}\prod_{i\neq j}\frac{\left[\sinh g(x_i-x_j)\right]\left[\sinh g(y_i-y_j)\right]}{g(x_i-x_j)g(y_i-y_j)} \times\notag\\
&\times\prod_{i,j}\left[2\cosh^{-1}g(x_i-y_j)\right]\left[2\cosh^{-1}\left(g(x_i-y_j)\right)\right]\,,
\end{align}
where $dX$ and $dY$ denote the opportune Haar measure, $g=k^{-\frac{1}{2}}$ and we neglect all the overall constants. Also, the Wilson loop insertion gets modified into
\begin{equation}
\langle W\rangle= \langle\sum_{i=1}^{N_1} e^{2gx_i}+\sum_{j=1}^{N_2}e^{2gy_j} \rangle\,.
\end{equation}
As explained in \cite{Gorini:2020new, Chester:2021gdw}, we can expand all the functions for $g\ll1$. The result can be expressed as a sum of Gaussian averages of matrix multitrace insertions
\begin{equation}
t_\nu(X)=\left(\Tr X\right)^{\nu_1}\left(\Tr X^2\right)^{\nu_2}\dots \left(\Tr X^k\right)^{\nu_k}\,,
\end{equation}
where $\nu$ is a collective notation for the set of integers $\nu_1, \,\nu_2,\dots, \nu_k$.
We have similar insertions also for $Y$. The explicit integral reads
\begin{equation}
\langle t_\nu(X) t_\lambda(Y) \rangle=\frac{\int dXdY\, e^{-g (X^2-Y^2)} t_\nu(X)t_\lambda(Y)}{\int dXdY\, e^{-g (X^2-Y^2)}}\,,
\end{equation}
where $\lambda$ is another arbitrary set of integers.
We compute the integrals using the results of \cite{Itzykson:1990zb}, recently reviewed in \cite{Chester:2021gdw}.
We push the expansion up to $k^{-3}$. We get
\begin{equation}\label{1ptres}
\langle\mathcal{O}(\tau)\rangle_W= \frac{N_1 N_2}{4k r\left( N_1 + N_2 \right)}-\frac{\pi ^2 N_1 N_2 \left(N_1 N_2-3\right)}{24 r\,k^3 \left(N_1+N_2\right) }+O(k^{-4})\,.
\end{equation}

With the same method, we calculate the two-point function. This is captured by the matrix model average
\begin{equation}
\pdv[2]{m}\langle W[m] \rangle=(\pi r)^2\langle \left(\sum_{i,j}\tanh(\pi(\lambda_i-\mu_j))\right)^2-\sum_{i,j}\frac{1}{\cosh^2(\pi\left(\lambda_i-\mu_j)\right)}\rangle_W\,.
\end{equation}
After similar computations we find
\begin{equation}\label{2ptres}
\langle\mathcal{O}(\tau_1)\mathcal{O}(\tau_2) \rangle_W =  \frac{N_1 N_2}{64 \pi ^2 r^2}-\frac{N_1 N_2 \left(N_1^2+N_2^2-14\right)}{384 k^2 r^2}+O(k^{-4})\,.
\end{equation}
One can check that the result differs from the two-point function without the Wilson loop already in the term proportional to $k^{-2}$ \cite{Gorini:2020new}. The reason is that the topological operators have a non-trivial bulk-to-boundary OPE with the defect.

In the following section, we perform some perturbative checks and elaborate on the relation with the stress tensor.

\section{Discussion and perturbative checks}\label{sec4}

As a further check of our result and to give also a somewhat different insight into defect correlation functions, we compute the first perturbative orders using Feynman diagrams.
We will expand the action and the Wilson loop for $k\gg1$ and evaluate all possible contractions among the local operators. 
We will limit to checking the leading order for the one-point function. For the two-point function, we verify the vanishing of the first quantum correction. We also discuss a general formula for the bremsstrahlung function in $\mathcal{N}=6$ SCFTs.

\subsection{One-point function and Bremmstrahlung} 

\begin{figure}
	\centering
	\subfigure[]{\begin{tikzpicture} \begin{feynman}
				\vertex (a); 
				\vertex[right=2 cm of a] (b) ;
				\vertex[right=2 cm of b] (c) ; 
				\vertex[above=2cm of b] (d) ;
				\diagram* {(b) -- [scalar,half left] (d), (b) -- [scalar, half right] (d),};
				\draw[blue,thick] (a) -- (c);
				\draw[fill= gray] (d) circle (3pt);
				\draw[fill=white] (b) circle (2pt);
		\end{feynman} \end{tikzpicture} \label{1pt:feynmana}}
			\subfigure[]{\begin{tikzpicture} \begin{feynman}
				\vertex (a); 
				\vertex[right=2 cm of a] (b) ;
				\vertex[right=2 cm of b] (c) ; 
				\vertex[above=1cm of b] (d) ;
				\vertex[above=1cm of d] (e) ;
				\diagram* {(b) -- [gluon] (d),};
				\diagram* {(d) -- [scalar,half left] (e), (d) -- [scalar, half right] (e),};
				\draw[blue,thick] (a) -- (c);
				\draw[fill=gray] (e) circle (3pt);
				\draw[fill=white] (b) circle (2pt);
		\end{feynman} \end{tikzpicture}\label{1pt:feynmanb}}
				\caption{Feynman diagrams for the leading contribution of the defect one-point function.}	\label{1pt:feynman}

\end{figure}
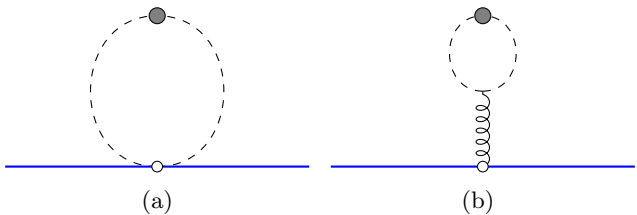

We  compute $\langle \mathcal{O}(\tau)\rangle_W$ in perturbation theory.
Since non-zero diagrams require interaction with the Wilson loop, the first contribution is of order $k^{-1}$. At this order in $k$, the Feynman diagrams are those in figure \ref{1pt:feynman}.

It is easy to evaluate the contribution of the diagram \ref{1pt:feynmanb}.
After the Wick contractions, the diagram is proportional to $U^I\bar V_I\equiv0$. Then, we conclude that the contribution of the diagram \ref{1pt:feynmanb} is zero\footnote{One can also check that the integral is vanishing for parity.}.

We turn to the diagram \ref{1pt:feynmana}, and we find 
\begin{align}
\langle W\mathcal{O}(0)\rangle&=\frac{2}{N_1+N_2}\Tr(U^A C_A \bar V_B \bar C^B)(-i)\left(-\frac{2\pi i}{k}\right){M_I}^J\int_{-\infty}^\infty ds\, \Tr(\bar C^IC_J(s))=\notag\\
&=-\frac{1}{16 \pi^2}\frac{4\pi}{k}\frac{N_1 N_2}{N_1+N_2}\frac{U^I{M_I}^J\bar V_J}{r}\int_{-\infty}^\infty ds\, \frac{1}{1+s^2}=\frac{1}{4kr}\frac{N_1 N_2}{N_1+N_2}\,,
\end{align}
where we used that $U^I(\tau){M_I}^J\bar V_J(\tau)\equiv-\frac{1}{2}$. The result is in perfect agreement with the matrix model prediction of Eq.~\eqref{1ptres}.

We now give the precise physical interpretation for this one-point function. Since $\mathcal{O}(0)$ is the twisted operator of the superconformal primary ${O_I}^J$, which is part of the stress tensor multiplet, there must be a supersymmetric Ward identity relating $\langle\mathcal{O}(0)\rangle_W$ and $\langle T_{\mu\nu}\rangle_W$. As explained in Sec.~\ref{sec1.3}, all the physical information of $\langle T_{\mu\nu}\rangle_W$ is contained in the constant $h_T$ of Eq.~\eqref{ht}. Then, using conformal symmetry, we conclude that 
\begin{equation}
\langle\mathcal{O}(0)\rangle_W\propto\frac{h_t}{r}\,.
\end{equation}
The quantity $h_T$ is in turn related to the bremsstrahlung function $B$ defined by Eq.~\eqref{eq4.5} by the Ward identity $B=2h_T$ \cite{Lewkowycz:2013laa}.
The explicit formula was first conjectured by separating the radiation component from the self-energy part of the field. Then, the relation was shown to hold in full generality by using dCFT considerations \cite{Bianchi:2018zpb, Bianchi:2019sxz}. In the case $N_1=N_2$, the bremsstrahlung is known, and we can compare it with our result to fix the missing constant of proportionality. In the end, we find \footnote{It would be interesting to derive the result more formally from a Ward identity like in \cite{Fiol:2015spa}, perhaps by using the superspace introduced in \cite{Liendo:2015cgi}. } \cite{Griguolo:2012iq, Lewkowycz:2013laa, Bianchi:2014laa, Bianchi:2017svd, Bianchi:2017ozk, Bianchi:2018bke, Bianchi:2018scb}
\begin{equation}
\langle\mathcal{O}(0)\rangle_W=\frac{2h_T}{r}=\frac{B}{r}\,.
\end{equation}
However, this conclusion depends only on the structure of the $\mathcal{N}=6$ stress tensor multiplet and not on the specific theory and therefore holds in any $\mathcal{N}=6$ SCFT. 
Then, we can read the bremsstrahlung function for the fermionic Wilson line in the $U(N_1)_k\times U(N_2)_{-k}$ model 
\begin{equation}
B(k,N_1,N_2)= \frac{  N_1 N_2 }{4k(N_1+N_2)}-\frac{\left(\pi ^2 N_1 N_2 \left(N_1 N_2-3\right)\right) }{24 k^3\left(N_1+N_2\right)}+O(k^{-4})\,.
\end{equation}
This formula was already conjectured by extending a prescription, valid in the limit $N_1=N_2$, based on a worldvolume deformation of the BPS $\frac{1}{2}$--BPS circular Wilson loop known as latitude \cite{Bianchi:2017ozk}. 
However, to the best of our knowledge, a rigorous argument was still missing.

Combining all these considerations, we propose a general formula for any BPS superconformal line $\mathcal{D}$ preserving at least $\mathfrak{su}(2)$ R-symmetry factor in any $\mathcal{N}=6$ SCFT
\begin{equation}
B= \frac{i}{8\pi^2 r\langle\mathcal{D}\rangle} \frac{\partial\langle \mathcal{D}\rangle[m]}{\partial m}\bigg|_{m = 0}\,.
\end{equation}
According to the general classification of \cite{Agmon:2020pde}, these are $\frac{1}{3}$--BPS lines (see \cite{Drukker:2022txy} for a recent explicit example). We stress again that there is no dependence on the specific realization of theory.
For SCFTs with $\mathcal{N}=4,5$ supersymmetry, the first derivative of the vev of the defect is no longer related to the one-point function of the stress tensor but rather to that of flavor currents.

\subsection{The two-point function}

We now move to the two-point function.
The computation of the tree-level is straightforward\footnote{We are using a shorthand notation $\tau_i\to(r\cos\tau_i,r\sin\tau_i,0)$, $i=1,2$.}
\begin{equation}
\langle \mathcal{O}(\tau_1)\mathcal{O}(\tau_2)\rangle_W^{(0)}=N_1N_2\frac{U(\tau_1)\cdot \bar V(\tau_2)U(\tau_2)\cdot \bar V(\tau_1)}{16\pi^2 |\tau_1-\tau_2|}=\frac{N_1N_2}{r^264\pi^2}\,.
\end{equation}
As a consistency check, we push the perturbative computation to 1-loop. The diagrams contributing to this order are those in figure \ref{2pt:1loop}.

\begin{figure}
	\centering
	\subfigure[]{\begin{tikzpicture} \begin{feynman}
				\vertex (a); 
				\vertex[right=1. cm of a] (b) ;
				\vertex[right=2 cm of b] (c) ; 
				\vertex[right=1.cm of c] (d) ;
				\vertex[above=2cm of b] (e) ;
				\vertex[above=2cm of c] (f) ;
				\vertex[right=2cm of a] (h) ;
				\vertex[above=1.125cm of h] (g) ;
				\vertex[above=2.87cm of h] (l) ;
				\diagram* {(e) -- [scalar,half left] (f), (e) -- [scalar, half right] (f),};
				\diagram* {(g) -- [gluon] (l)};
				\draw[blue,thick] (a) -- (d);
				\draw[fill= gray] (e) circle (3pt);
				\draw[fill=gray] (f) circle (3pt);
		\end{feynman} \end{tikzpicture}\label{2pt:1loopc}}
			\subfigure[]{\begin{tikzpicture} \begin{feynman}
				\vertex (a); 
				\vertex[right=1 cm of a] (b) ;
				\vertex[right=1 cm of b] (c) ; 
				\vertex[right=1 cm of c] (k) ; 
				\vertex[right=1cm of k] (d) ;
				\vertex[above=2cm of b] (e) ;
				\vertex[above=2cm of k] (f) ;
				\vertex[right=2cm of a] (h) ;
				\diagram* {(e) -- [scalar] (f), (e) -- [scalar] (c),(f) -- [scalar] (c),};
				\draw[blue,thick] (a) -- (d);
				\draw[fill= gray] (e) circle (3pt);
				\draw[fill=gray] (f) circle (3pt);
				\draw[fill=white] (c) circle (2pt);
		\end{feynman} \end{tikzpicture}\label{2pt:1loopd}}
					\subfigure[]{\begin{tikzpicture} \begin{feynman}
				\vertex (a); 
				\vertex[right=1 cm of a] (b) ;
				\vertex[right=1 cm of b] (c) ; 
				\vertex[right=1 cm of c] (k) ; 
				\vertex[right=1cm of k] (d) ;
				\vertex[above=2cm of b] (e) ;
				\vertex[above=2cm of k] (f) ;
				\vertex[right=2cm of a] (h) ;
				\vertex[above=1.12 cm of c] (u) ; 
				\diagram* {(e) -- [scalar,half left] (f), (e) -- [scalar, half right] (f),};
				\diagram* {(c) -- [gluon] (u)};
				\draw[blue,thick] (a) -- (d);
				\draw[fill=gray] (f) circle (3pt);
				\draw[fill= gray] (e) circle (3pt);
				\draw[fill=gray] (f) circle (3pt);
				\draw[fill=white] (c) circle (2pt);
		\end{feynman} \end{tikzpicture}\label{2pt:1loope}}
							\subfigure[]{\begin{tikzpicture} \begin{feynman}
				\vertex (a); 
				\vertex[right=0.5 cm of a] (u) ;
				\vertex[right=3 cm of a] (n) ;
				\vertex[right=1 cm of a] (m) ;
				\vertex[right=2 cm of a] (c) ; 
				\vertex[right=3.5 cm of a] (v) ;
				\vertex[right=4cm of a] (d) ;
				\vertex[above=2cm of m] (e) ;
				\vertex[above=2cm of n] (f) ;
				\diagram* {(m) -- [fermion,half left] (n),};
				\diagram* {(e) -- [scalar,half left] (f), (e) -- [scalar, half right] (f),};
				\draw[blue,thick] (a) -- (d);
				\draw[fill=gray] (f) circle (3pt);
				\draw[fill= gray] (e) circle (3pt);
				\draw[fill=white] (m) circle (2pt);
				\draw[fill=white] (n) circle (2pt);
		\end{feynman} \end{tikzpicture}\label{2pt:1loopf}}
		\caption{Feynman diagrams for the 1-loop correction of the defect two-point function.}\label{2pt:1loop}
\end{figure}
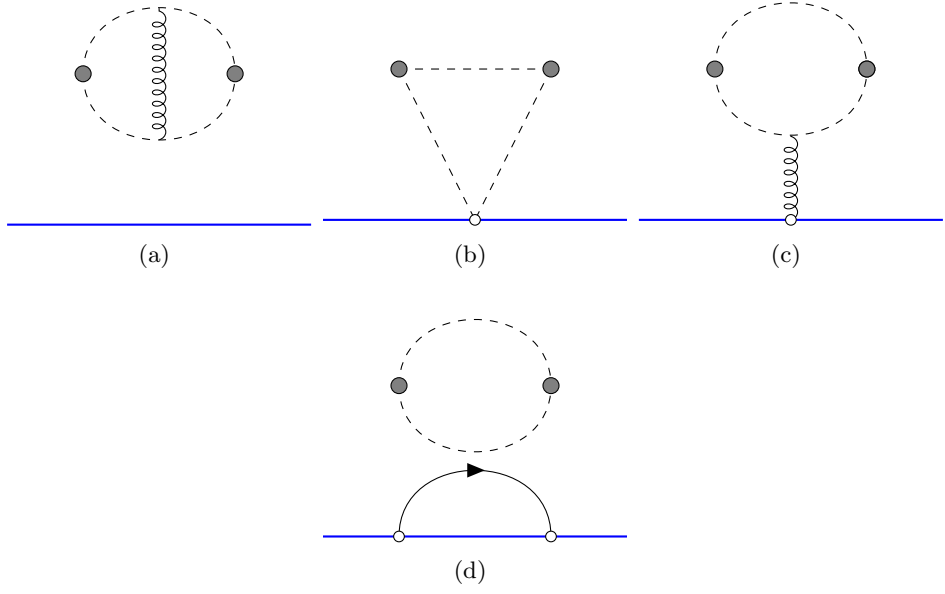

\begin{figure}
	\centering
	{\begin{tikzpicture} \begin{feynman}
				\vertex (a); 
				\vertex[right=0.5 cm of a] (u) ;
				\vertex[right=3 cm of a] (n) ;
				\vertex[right=1 cm of a] (m) ;
				\vertex[right=2 cm of a] (c) ; 
				\vertex[right=3.5 cm of a] (v) ;
				\vertex[right=4cm of a] (d) ;
				\vertex[above=2cm of m] (e) ;
				\vertex[above=2cm of n] (f) ;
				\diagram* {(m) -- [fermion,half left] (n),};
				\draw[blue,thick] (a) -- (d);
				\draw[fill=white] (m) circle (2pt);
				\draw[fill=white] (n) circle (2pt);
		\end{feynman} \end{tikzpicture}}
				\caption{The Feynman diagram for the 1-loop correction of vev of the fermionic Wilson loop.}\label{wl:1loop}
\end{figure}
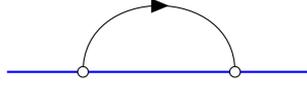

We see that in the diagrams \ref{2pt:1loopc}, \ref{2pt:1loopd}, and \ref{2pt:1loope}, the operators interact with the Wilson loop, while the diagram \ref{2pt:1loopf} is a vacuum correction. Therefore, \ref{2pt:1loopf} is canceled against the 1-loop correction $\delta W$ to the vev of the Wilson loop of figure \ref{wl:1loop} coming from the normalization. Indeed, since\footnote{In the equations, we denote the contribution from a given Feynman diagram with the number and the letter associated with the corresponding figure.}

\begin{equation}
\mathrm{\ref{2pt:1loopf}}= \langle \mathcal{O}(\tau_1)\mathcal{O}(\tau_2)\rangle_W^{(0)}\delta W\,,
\end{equation}
we have
\begin{align}
\langle \mathcal{O}(\tau_1)\mathcal{O}(\tau_2)\rangle_W&=\frac{\langle \mathcal{O}(\tau_1)\mathcal{O}(\tau_2)\rangle^{(0)}+\langle \mathcal{O}(\tau_1)\mathcal{O}(\tau_2)\rangle^{(1)}}{1+\delta W}=\\
&=\langle \mathcal{O}(\tau_1)\mathcal{O}(\tau_2)\rangle^{(0)}+\mathrm{\ref{2pt:1loopc}}+\mathrm{\ref{2pt:1loopd}}+\mathrm{\ref{2pt:1loope}}\notag\,.
\end{align}
The term $\delta W$ is both IR and UV divergent and can be found in \cite{Griguolo:2012iq}, so its cancellation is crucial to ensure supersymmetry. Let us discuss the remaining contributions.

As explained in \cite{Gorini:2020new} on the line, the diagram \ref{2pt:1loopc} is vanishing for the kinematics of the Chern-Simons propagator, which always leads to an integrand function odd in at least one of the loop integration variables.

Then we write down the diagram \ref{2pt:1loopd}, and we perform the Wick contractions
\begin{align}
\mathrm{\ref{2pt:1loopd}}=&-\frac{4\pi }{k}\Tr(U^A C_A \bar V_B \bar C^B)(\tau_1)\Tr(U^K C_K \bar V_L \bar C^L)(\tau_2){M_I}^J\int_{-\infty}^\infty ds\, \Tr(\bar C^IC_J(s))=\notag\\
=&-\frac{1}{16 \pi k |\tau_1-\tau_2|}\frac{N_1 N_2}{N_1+N_2}
\big(U^I(\tau_1){M_I}^J\bar V_J(\tau_2) U^K(\tau_2) V_K(\tau_1)+ \notag\\
&U^I(\tau_1){M_I}^J\bar V_J(\tau_2) U^K(\tau_2) V_K(\tau_1)\big)\equiv 0\,,
\end{align}
after the evaluation of the contraction of the polarization vector with ${M_I}^J$.

Finally, we consider the diagram \ref{2pt:1loope}. We simplify the computation using the topologicity of the operators. 
Indeed, if we set $\tau_1=\pi/2$ and $\tau_2=3\pi/2$, we get again the integral over the entire spacetime of an odd function, which is zero.
In the end, we find that there is no 1-loop correction, as in Eq.~\eqref{2ptres}. This simple result holds only for topological operators in this specific kinematical configuration. It is not necessarily true for any two-point function of the superconformal primary of the stress tensor multiplet.

\section{Conclusion and outlook}

In this work, we have constructed and computed supersymmetric correlation functions involving both local operators and line operators in 3d $\mathcal{N}=4$ theories.
The local operators are those forming the topological sector \cite{Chester:2014mea}, which has been shown to be compatible with BPS line defects. We have described this setup for $\frac{1}{2}$--BPS line defects both in UV $\mathcal{N}=4$ super Yang-Mills theory on $S^3$ and $\mathcal{N}=4$ SCFTs. For dimension one operators, we have argued that the vev of the loop operator with specific supersymmetric flavor deformations is the generating functional for integrated defect correlation functions.
Once the vev of the defect is known, for example by localization, defect correlation functions can be computed.

We have applied this method to the $\frac{1}{2}$--BPS Wilson line in the $U(N_1)_k\times U(N_2)_{-k}$ ABJM model and computed the one- and two-point functions of the topological operators of dimension 1 for the Chern-Simons level $k\gg 1$. In this limit, the theory becomes perturbative, allowing for an explicit check of our results with an honest Feynman diagram calculation.
Since this operator is part of the stress tensor multiplet, we managed to relate the one-point function to the so-called bremsstrahlung function, whose expression was rigorously proved only for the case $N_1=N_2$. Comparing our perturbative result with those available in the literature \cite{Griguolo:2012iq, Lewkowycz:2013laa, Bianchi:2014laa, Bianchi:2017svd, Bianchi:2017ozk, Bianchi:2018bke, Bianchi:2018scb, Griguolo:2021rke}, we infer an exact formula for the bremsstrahlung valid for all $\frac{1}{3}$--BPS lines in $\mathcal{N}=6$ SCFT, regardless of the specific details of the models. We also present the first result for a two-point function in ABJM with BPS line defects. That is the simplest observable exhibiting crossing symmetry. We expect our results to be relevant to extract CFT data and as a crosscheck with other methods, such as bootstrap and integrability.

There are several directions to explore in the future. While there are many examples of defect correlation functions in 4d SCFTs in different regimes and with different amounts of supersymmetry, the situation in 3d is still rather understudied.
The first possibility is to improve the matrix model computation. A natural method for doing that is the Fermi gas technique \cite{Marino:2011eh}, which has already been developed for BPS Wilson lines \cite{Klemm:2012ii} and mass deformations \cite{Nosaka:2015iiw}, but not for mass deformations with the Wilson loop \footnote{See also \cite{Chester:2020jay, Gaiotto:2020vqj, Hatsuda:2021oxa} for applications to the topological sector without defects.}.
The combination with the technology developed in this paper would allow us to access the defect correlation function in the strongly coupled regime and compare it with the string or M-theory dual. 

It would be interesting to analyze our defect correlation functions from the bootstrap perspective. Without defects, the topological sector is captured by a consistent and much simpler truncation of the full bootstrap equation. It turns out that the operator algebra is captured by a specific algebraic structure, namely a deformation quantization of the Higgs (or Coulomb) branch chiral ring \cite{Beem:2016cbd}. It would be interesting to understand how this picture is modified by the presence of the defect, and to study the corresponding problem along the lines of \cite{Chang:2019dzt, Fan:2019jii}. In a similar direction, one could try to extend the IR formula of \cite{Gaiotto:2019mmf, Bullimore:2020jdq} for topological correlation functions\footnote{We thank D. Gaiotto and M. Bullimore for suggesting this possibility.}.

Moreover, the protected sector can provide useful information for the general defect bootstrap problem. While the study of defect CFT for the fermionic Wilson line has already begun \cite{Bianchi:2017ozk, Bianchi:2020hsz, Gorini:2022jws}, nothing is known for correlation functions of bulk operators with defects. We expect the existence of superconformal Ward identities, similarly to other supersymmetric systems in 4d \cite{Liendo:2016ymz, Liendo:2018ukf} and 6d \cite{Meneghelli:2022gps}. Perhaps it is also possible to use an inversion formula \cite{Caron-Huot:2017vep} as in \cite{Barrat:2021yvp}.

Finally, it would be interesting to study these observables for other defects, such as vortex loops \cite{Drukker:2008zx, Kapustin:2012iw, Drukker:2012sr}. One possible difficulty is the lack of a general localization scheme for these disorder operators in Chern-Simons matter theory. 

It would also be interesting to study the same problems in $\mathcal{N}=4$ QFTs, where derivatives of the deformed vev of the defect compute correlation functions of flavor current multiplets (see \cite{Chang:2019dzt} for the corresponding problem without defects).

\acknowledgments

It is a pleasure to thank Luca Griguolo, Domenico Seminara, Itamar Yaakov, and Stefano Cremonesi for interesting discussions and useful insights. We are especially grateful to Luca Griguolo and Itamar Yaakov for useful comments on the draft.
We also thank the Department of Mathematical Sciences at Durham University and its faculty members for hospitality, discussions, and feedback when presenting this work in the local journal club.
This work has been supported in part by Della Riccia Foundation, Italian Ministero dell'Universit\`a e Ricerca (MUR), and by Istituto Nazionale di Fisica Nucleare (INFN) through the ``Gauge Theories, Strings, Supergravity'' (GSS) and ``Gauge and String Theory'' (GAST) research projects.

\appendix

\section{ABJ(M) action and Feynman rules}\label{ABJM}\label{appa}

Here we briefly summarize the basic notation about ABJM theory that will be needed in this paper.
We work in Euclidean space with coordinates $x^\mu=(x^1,x^2,x^3)$ and metric $\delta_{\mu\nu}$. 
We take the 3d flat Clifford algebra $\gamma^\mu$ to be generated by the Pauli matrices $(\gamma^\mu)_\alpha^{\; \beta} \equiv (\sigma^\mu)_\alpha^{\; \beta}$, $\mu = 1,2,3$.
Spinorial indices are raised and lowered according to
\[
\psi^\alpha=\varepsilon^{\alpha\beta}\psi_\beta,\qquad \psi_\alpha=\varepsilon_{\alpha\beta}\psi^\beta\,, \qquad {\rm with } \qquad \varepsilon^{12}=-\varepsilon_{12}=1\,.
\]
\vskip 10pt

The field content of the $U(N_1)_k \times U(N_2)_{-k}$ ABJ(M) theory includes two gauge fields $(A_\mu)_i^j$, $(\hat A_\mu)_{\hat i}^{\hat j}$ belonging to the adjoint representation of $U(N_1)$ and $U(N_2)$ respectively, minimally coupled to four matter multiplets $(C_I, \bar{\psi}^I)_{I=1, \dots , 4}$ in the $(N_1, \bar{N}_2)$ representation of the gauge group and their conjugates $(\bar{C}^I, \psi_I)_{I=1, \dots , 4}$ in the $(\bar{N}_1, N_2)$. 

The Euclidean action is given by 
\begin{equation}\label{action}
S=S_{\textup{CS}}+S_{\textup{mat}} + S_{\textup{pot}}\,,
\end{equation}
where $S_{\textup{CS}}$ contain the Chern-Simons kinetic terms, $S_{\textup{mat}}$ the kinetic terms for the matter fields and the gauge-matter interactions, and $S_{\textup{pot}}$ the Yukawa, and the potential terms. For the explicit expressions, we refer to \cite{Bianchi:2018bke,Gorini:2020new, Gorini:2022jws}.

The action is invariant under the $\mathcal{N}=6$ superconformal algebra $\mathfrak{osp}(6|4)$. Its explicit realization is
\bea \label{susytransf}
\delta C_K&=&- \bar\xi^{IJ,\alpha}\ \varepsilon_{IJKL}\ \bar \psi^{L}_\alpha \non\,,  \\
\delta\bar C^K&=&2\bar\xi^{KL,\alpha}\ \psi_{L,\alpha} \non \,,\\
\delta\bar\psi^{K,\beta}&=& 2i\bar\xi^{KL,\alpha}{(\gamma^\mu)_\alpha}^\beta D_\mu C_L + \frac{4\pi i}{k}\bar\xi^{KL,\beta}(C_L\bar C^M C_M-C_M\bar C^MC_L) 
+ \frac{8\pi i}{k}\bar\xi^{IJ,\beta}C_I\bar C^KC_J \non \\
&&\quad - \, 2i\bar\eta^{KL,\beta} C_L \notag \non\,, \\
\delta\psi^\beta_K&=&-i\bar\xi^{IJ,\alpha}\varepsilon_{IJKL}{(\gamma^\mu)_\alpha}^\beta D_\mu \bar C^L+\frac{2\pi i}{k}\bar\xi^{IJ,\beta}\varepsilon_{IJKL}(\bar C^LC_M\bar C^M-\bar C^MC_M\bar C^L)\non \\
&&\quad +\frac{4\pi  i}{k}\bar\xi^{IJ,\beta}\varepsilon_{IJML}\bar C^M C_K\bar C^L+i\bar\eta^{IJ,\beta}\varepsilon_{IJKL}\bar C^L\non\,, \\
\delta A_\mu&=&\frac{4\pi i}{k}\bar\xi^{IJ, \alpha}{(\gamma_\mu)_\alpha}^\beta\bigg(C_I\psi_{J\beta} - \frac{1}{2}\varepsilon_{IJKL}\bar\psi^K_\beta\bar C^L\bigg) \non\,, \\
\delta\hat A_\mu&=&\frac{4\pi i}{k}\bar\xi^{IJ, \alpha}{(\gamma_\mu)_\alpha}^\beta\bigg(\psi_{J\beta}C_I -\frac{1}{2}\varepsilon_{IJKL}\bar C^L\bar\psi^K_\beta\bigg)\,,
\eea
where the parameters of the transformations are expressed in terms of conformal Killing spinors, whose flat space expression is
\beq\label{flatks}
\bar\xi^{IJ}_{\alpha} = \bar\theta^{IJ}_{\alpha} - x^\mu (\gamma_\mu)_\a^{\; \b} {\bar \eta}^{IJ}_\b\,.
\eeq

For perturbative computations, we also need the tree-level propagators.
After rescaling the gauge fields in the action read as 
\beq\label{eq:rescaling}
A_\mu \to \frac{1}{\sqrt{k}} A_\mu \,,\qquad  \qquad \hat{A}_\mu \to \frac{1}{\sqrt{k}} \hat{A}_\mu\,,
\eeq
they read
\begin{itemize}
	\item Scalar propagator
	\begin{align}
	\langle {(C_I)_i}^{\hat{j}} (x)\ {(\bar C^J)_{\hat k}}^l (y)  \rangle &=\delta^J_I\delta^l_i\delta^{\hat j}_{\hat k} \ \frac{\Gamma(\frac{1}{2}-\epsilon)}{{4\pi}^{\frac{3}{2}-\epsilon}}\frac{1}{{|x-y|}^{1-2\epsilon}} \,.\label{scalartree}\\
	\end{align}  

	\item Fermion propagator
	\begin{equation}
	\label{0fermion}
	\langle {(\psi_{\alpha I})_{\hat i}}^j (x)\ {(\bar \psi^{J\beta})^{\hat l}}_k (y)  \rangle= \delta^J_I\delta^{\hat l}_{\hat i}\delta^j_k \ i \, \frac{\Gamma(\frac{3}{2}-\epsilon)}{{2\pi}^{\frac{3}{2}-\epsilon}} \ {(\gamma^\mu)_\alpha}^\beta\ \frac{(x-y)_\mu}{{|x-y|}^{3-2\epsilon}}\,.
	\end{equation}

	\item Vector propagators in Landau gauge 
\bea\label{prop:vector}
	&& \langle {(A_\mu)_i}^j (x)\ {(A_\nu)_k}^l (y)  \rangle =  \delta^l_i\delta^j_k \   i  \, \frac{\Gamma(\frac{3}{2}-\epsilon)}{{\pi}^{\frac{1}{2}-\epsilon}}\ \varepsilon_{\mu\nu\rho}\ \frac{(x-y)^\rho}{{|x-y|}^{3-2\epsilon}} \non\,, \\
	&&	\langle {(\hat A_\mu)_{\hat i}}^{\hat j} (x)\ {(\hat A_\nu)_{\hat k}}^{\hat l} (y)  \rangle =  - \, \delta^{\hat l}_{\hat i}\delta^{\hat j}_{\hat k} \   i  \, \frac{\Gamma(\frac{3}{2}-\epsilon)}{{\pi}^{\frac{1}{2}-\epsilon}}\ \varepsilon_{\mu\nu\rho}\ \frac{(x-y)^\rho}{{|x-y|}^{3-2\epsilon}}\,.
\eea

\end{itemize}

\section{From $\mathcal{N}=6$ to $\mathcal{N}=4$ superalgebras} \label{appb}

In this appendix, we spell out the details of the supersymmetry algebra that appears in this paper. As in \cite{Beem:2013sza}, we work with complexified algebras.

\paragraph{$\mathcal{N}=6$ superalgebra}
$\mathcal{N}=6$ SCFTs in 3d are invariant under the $\mathfrak{osp}(6|4)$ superalgebra. It contains the usual 3d conformal algebra, whose commutation relations are
\begin{equation}
\begin{aligned}\label{balgebra}
[M^{\mu\nu}, M^{\rho\sigma}]&=\delta^{\sigma\mu}M^{\nu\rho}-\delta^{\sigma\nu}M^{\mu\rho}+\delta^{\rho\nu}M^{\mu\sigma}-\delta^{\rho\mu}M^{\nu\sigma}\,, &[P^\mu,K^\nu]&=2(\delta^{\mu\nu}D+M^{\mu\nu})\,, \\
[P^\mu,M^{\nu\rho}]&=\delta^{\mu\nu}P^{\rho}-\delta^{\mu\rho}P^\nu\,, &[K^\mu,M^{\nu\rho}]&=\delta^{\mu\nu}K^{\rho}-\delta^{\mu\rho}K^\nu\,,\\
[D,P^\mu]&=P^\mu\,, &[D,K^\mu]&=-K^\mu\,.
\end{aligned}
\end{equation}
The spacetime generators act on scalar operators as
\begin{subequations}
\begin{align}
\comm{P_\mu}{O(x)}&=-i\partial_\mu O(x)\,,\\
\comm{K_\mu}{O(x)}&=i(-x^2\partial_\mu+2x_\mu x\cdot\partial+2x_\mu\Delta) O(x)\,,\\
\comm{M_{\mu\nu}}{O(x)}&=(x_\mu\partial_\nu-x_\nu\partial_\mu) O(x)\,,\\
\comm{D}{O(x)}&=(-x^\mu\partial_\mu-\Delta) O(x)\,,
\end{align}
\end{subequations}
where $\Delta$ is the dimension of the operator $O(x)$.
The bosonic part of the algebra contains also an $\mathfrak{su}(4)\simeq\mathfrak{so}(6)$ R-symmetry factor, generated by the traceless matrices ${J_I}^J$, with $I,J=1,\dots,4$. Their algebra reads
\begin{equation}\label{Ralgebra}
[{J_I}^J, {J_K}^L]=\delta^L_I {J_K}^J-\delta^J_K {J_I}^L\,.
\end{equation}
We have defined implicitly also the action of ${J_I}^J$ on the (anti-)fundamental representation. That is, a generic operator $\Phi_I$ ($\bar{\Phi}^I$) transforms under ${J_I}^J$ according to
\begin{equation}\label{fund4}
[{J_I}^J, \Phi_K] =\frac{1}{4}\delta^J_I \Phi_K-\delta^J_K \Phi_I \,,\qquad \qquad 
[{J_I}^J, \bar{\Phi}^K]=\delta^K_I \bar{\Phi}^J - \frac{1}{4}\delta^J_I \bar{\Phi}^K\,.
\end{equation}

The odd generators are $Q^{IJ}_\alpha$, $S^{IJ}_\alpha$ and close the following algebra
\begin{equation}\label{QSalgebra}
\begin{aligned}
\{Q^{IJ}_\alpha, Q^{KL,\beta}\}&= \varepsilon^{IJKL}{(\gamma^\mu)_{\alpha}}^\beta P_\mu\,,\qquad\qquad \{S_\alpha^{IJ}, S^{\beta KL}\}= \varepsilon^{IJKL}{(\gamma^\mu)_{\alpha}}^\beta K_\mu\,,\\
\{Q^{IJ}_\alpha, S^{\beta KL}\}&= \varepsilon^{IJKL}\left(\frac12 {(\gamma^{\mu\nu})_\alpha}^\beta M_{\mu\nu}+\delta^\beta_\alpha D\right)+\delta^\beta_\alpha \varepsilon^{KLMN}(\delta^J_M {J_N}^I-\delta^I_M {J_N}^J)\,,   
\end{aligned}
\end{equation}
and similarly for $\bar Q_{\alpha IJ}=\frac{1}{2}\varepsilon_{IJKL}Q^{KL}_\alpha$ and $\bar S_{\alpha IJ}=\frac{1}{2}\varepsilon_{IJKL}S^{KL}_\alpha$. 
Finally, the mixed commutators are
\begin{equation}
\begin{aligned}
[K^\mu,Q_{\alpha}^{IJ}]&={(\gamma^\mu)_\alpha}^\beta S^{ IJ}_\beta\,, &[P^\mu,S_{\alpha}^{IJ}]&={(\gamma^\mu)_\alpha}^\beta Q^{ IJ}_\beta\,,\\
[M^{\mu\nu}, Q_{\alpha}^{IJ}]&=-\frac{1}{2}{(\gamma^{\mu\nu})_\alpha}^\beta Q_{\beta}^{IJ}\,, &
[M^{\mu\nu}, S_{\alpha}^{IJ}]&=-\frac{1}{2}{(\gamma^{\mu\nu})_\alpha}^\beta S_{\beta}^{IJ}\,,\\
[D, Q_{\alpha}^{IJ}]&=\frac{1}{2}Q_{\alpha}^{IJ}\,, &[D, S^{\alpha IJ}]&=-\frac{1}{2}S^{\alpha IJ} \,,\\
[{J_I}^J, Q^{KL}_\alpha]&=\delta^K_I Q^{JL}_\alpha+\delta^L_I Q^{KJ}_\alpha-\frac{1}{2}\delta^J_IQ^{KL}_\alpha\,, &[{J_I}^J, S^{\alpha KL}]&=\delta^K_I S^{\alpha JL}+\delta^L_I S^{\alpha KJ}-\frac{1}{2}\delta^J_I S^{\alpha KL}\,.
\end{aligned}
\end{equation}
The definition of $\delta$ in terms of $\bar Q_{\alpha IJ}$ and $\bar S_{\alpha IJ}$ is
\begin{equation}
\delta\Phi=\comm*{\bar\theta^{IJ }\bar Q_{IJ}-i\bar\eta^{IJ} \bar S_{IJ}}{\Phi}\,.
\end{equation}


\paragraph{$\frac{1}{2}$--BPS Wilson line in ABJM}

We briefly describe the superalgebra $\mathfrak{su}(1,1|3)\oplus\mathfrak{u}(1)_b$ preserved by the fermionic Wilson loop. 
The maximal bosonic subalgebra of $\mathfrak{su}(1,1|3)$ is $\mathfrak{sl}(2) \oplus \mathfrak{su}(3) \oplus \mathfrak{u}(1)$, where $\mathfrak{sl}(2)\simeq\mathfrak{su}(1,1)$ is the Euclidean conformal algebra in one dimension and $\mathfrak{su}(3) \oplus \mathfrak{u}(1)$ is the R-symmetry algebra. The $\mathfrak{u}(1)_b$ factor is generated by $B=M_{12}+2i{J_1}^1$.

The  $\mathfrak{su}(1,1)$ algebra is generated by $P_3$,  $K_3$, and $D$. The $\mathfrak{su}(3)$ R-symmetry subalgebra is generated by traceless operators ${L_a}^b$, whose explicit form reads
\begin{equation}\label{Rgenerators}
{L_a}^b=
\renewcommand\arraystretch{1.2}\begin{pmatrix}
{J_2}^2+\frac{1}{3}{J_1}^1 & {J_2}^3 & {J_2}^4 \\
{J_3}^2 & {J_3}^3+\frac{1}{3}{J_1}^1 &  {J_3}^4 \\
{J_4}^2 &  {J_4}^3 &-{J_3}^3-{J_2}^2-\frac{2}{3}{J_1}^1  
\end{pmatrix}\,.
\end{equation}
These generators satisfy the algebraic relation
\begin{equation}\label{su3}
[{L_a}^b, {L_c}^d]=\delta_a^d {L_c}^b- \delta_c^b {L_a}^d   \,.  
\end{equation}
From Eq.~\eqref{fund4} and definitions \eqref{Rgenerators} it follows that the action of the $SU(3)$ R-symmetry generators on fields in the (anti-)fundamental representation is
\begin{equation}\label{fund3}
[{L_a}^b, \Phi_c]=\frac{1}{3}\delta^b_a \Phi_c-\delta^b_c \Phi_a\,, \qquad  [{L_a}^b, \bar{\Phi}^c]=\delta^c_a \bar{\Phi}^b - \frac{1}{3}\delta^b_a \bar{\Phi}^c\,.
\end{equation}
The spectrum of bosonic generators of $\mathfrak{su}(1,1|3)$ is completed by a residual $\mathfrak{u}(1)$ generator $M$, defined as
\begin{equation}\label{M}
M\equiv 3i M_{12}-2{J_1}^1    \,,
\end{equation}

We now move to the fermionic sector of the superalgebra. Since we have placed the line along the $x^3$-direction, the fermionic generators of the one-dimensional superconformal algebra are identified with the following supercharges
\begin{equation}
Q_1^{12}, Q_1^{13}, Q^{14}_1, Q_2^{23}, Q^{24}_2, Q^{34}_2\qquad\text{and}\qquad S_1^{12}, S_1^{13}, S^{14}_1, S_2^{23}, S^{24}_2, S^{34}_2\,.
\end{equation}
For an exhaustive description of the algebra, we refer to \cite{Gorini:2020new}.

\paragraph{$\mathcal{N}=4$ superalgebra}

In order to define a topological sector compatible with the Wilson line, we need to identify the $\mathcal{N}=4$ Poincar\'e subalgebra introduced in \cite{Dedushenko:2016jxl} in the ABJM superalgebra. In the following, we detail the precise decomposition.

To begin with, we choose an $\mathfrak{su}(2)_H\oplus\mathfrak{su}(2)_C$ R-symmetry algebra into $\mathfrak{su}(4)$.  
This can be explicitly realized by setting the $\mathfrak{su}(2)_H\oplus\mathfrak{su}(2)_C$ ${R_a}^b$, ${\bar{R}_{\dot a}}^{\;\;\dot b}$ as follows \footnote{The full pattern of symmetry breaking is $SU(4)\to SU(2)_H\times SU(2)_C\times U(1)_F$, where $U(1)_F$ is generated by $F={J_4}^4+{J_2}^2$. It commutes with the full $\mathfrak{osp}(4|4)$ superalgebra. Such generator will not play any role in our construction.}
\begin{align}
{R_a}^b=
\begin{pmatrix}
-\frac{{J_4}^4-{J_2}^2}{2} & {J_2}^4 \\ {J_4}^2 & \frac{{J_4}^4-{J_2}^2}{2}
\end{pmatrix}\,,\qquad
{\bar R_{\dot a}}^{\;\;\dot b}=
\begin{pmatrix}
-\frac{{J_3}^3-{J_1}^1}{2} & {J_1}^3 \\ {J_3}^1 & \frac{{J_3}^3-{J_1}^1}{2}
\end{pmatrix}\,.
\end{align}
The embedding makes $\mathfrak{su}(2)_H$ coincide with an $\mathfrak{su}(2)$ factor preserved by the Wilson line, namely the one spanned by ${L_1}^3$, ${L_3}^1$, ${L_1}^1-{L_3}^3$. Specifically, we choose the one used in the topological line of \cite{Gorini:2020new}.
The resulting algebra is 
\begin{equation}
\comm*{{R_a}^b}{{R_c}^d}=\delta_a^d {R_c}^b-\delta_c^b {R_a}^d\,,\qquad
\comm*{\bar R_{\dot a}^{\;\;\dot b}}{\bar R_{\dot a}^{\;\;\dot b}}=\delta_{\dot a}^{\dot d}\bar R_{\dot c}^{\;\;\dot b}-\delta_{\dot c}^{\dot b}\bar R_{\dot a}^{\;\;\dot d}\,.
\end{equation}
So far, we have broken the $\mathcal{N}=6$ R-symmetry group down to the $\mathcal{N}=4$. 
To identify the full $\mathfrak{osp}(4|4)$ algebra, we take the supercharges charged under our $\mathcal{N}=4$ R-symmetry subalgebra. They can be organized in the following way 
\begin{equation}
Q_{\alpha,a\dot a}=\begin{pmatrix} \bar Q_{\alpha,12} & \bar Q_{\alpha,32}\\ Q_{\alpha,14} & \bar Q_{\alpha,34}  \end{pmatrix}_{a\dot a}\,, \qquad
S_{\alpha,a\dot a}=\begin{pmatrix} \bar S_{\alpha,12} & \bar S_{\alpha,32}\\ S_{\alpha,14} & \bar S_{\alpha,34}  \end{pmatrix}_{a\dot a} \,.
\end{equation}
We find 
\begin{equation}
\comm*{{R_a}^b}{Q_{\alpha,c\dot a}}=\frac{1}{2}\delta_a^bQ_{\alpha,c\dot a}-\delta_c^bQ_{\alpha,a\dot a}\,,\qquad
\comm*{\bar R_{\dot a}^{\;\;\dot b}}{Q_{\alpha,a\dot c}}=\frac{1}{2}\delta_{\dot a}^{\dot b}Q_{\alpha,a\dot c}-\delta_{\dot c}^{\dot b}Q_{\alpha,a\dot a}\,.
\end{equation}
For an easier interpretation on $S^3$, we find it convenient to express the Poincar\'e generators using spinorial indices, and we shall use
\begin{align}
{P_\alpha}^\beta=&P^\mu{(\gamma_\mu)_\alpha}^\beta=\begin{pmatrix}P_3&P_1-iP_2\\P_1+iP_2 & -P_3 \end{pmatrix}\,,\\
{K_\alpha}^\beta=&K^\mu{(\gamma_\mu)_\alpha}^\beta=\begin{pmatrix}K_3&K_1-iK_2\\K_1+iK_2 & -K_3 \end{pmatrix}\,,\\
{M_\alpha}^\beta=&\frac{i}{2}\epsilon_{\mu\nu\lambda}M^{\mu\nu}{(\gamma_{\lambda})_\alpha}^\beta=\frac{1}{2}M^{\mu\nu}{(\gamma_{\mu\nu})_\alpha}^\beta=
\begin{pmatrix} iM_{12} & -M_{13}+iM_{23}\\M_{13}+iM_{23}& -iM_{12} \end{pmatrix}\,.
\end{align} 
Their algebra reads as
\begin{align}
\comm*{{P_\alpha}^\beta}{{K_\gamma}^\delta}&=2\left((2\delta_\alpha^\delta\delta_\gamma^\beta-\delta_\alpha^\beta\delta_\gamma^\delta)D+\delta^\beta_\gamma{M_\alpha}^\delta-\delta^\delta_\alpha{M_\gamma}^\beta \right)\,,\\
\comm*{{M_\alpha}^\beta}{{M_\gamma}^\delta}&=\delta_\gamma^\beta{M_\alpha}^\delta-\delta_\alpha^\delta{M_\gamma}^\beta\,,\\
\comm*{{M_\gamma}^\delta}{{P_\alpha}^\beta}&=\delta_\alpha^\beta{P_\gamma}^\delta+\delta_\gamma^\delta{P_\alpha}^\beta-2\delta_\gamma^\beta{P_\alpha}^\delta\,,\\
\comm*{{M_\gamma}^\delta}{{K_\alpha}^\beta}&=\delta_\alpha^\beta{K_\gamma}^\delta+\delta_\gamma^\delta{K_\alpha}^\beta-2\delta_\gamma^\beta{K_\alpha}^\delta\,.
\end{align}
Then, we can finally write down the $\mathcal{N}=4$ algebra.
The even-odd part reads
\begin{align}
\comm*{{P_\alpha}^\beta}{S_{\gamma,a\dot a}}&=2\delta_\gamma^\beta Q_{\alpha,a\dot a}-\delta_\alpha^\beta Q_{\gamma,a\dot a} \,,
&\comm*{{K_\alpha}^\beta}{Q_{\gamma,a\dot a}}&=2\delta_\gamma^\beta S_{\alpha,a\dot a}-\delta_\alpha^\beta S_{\gamma,a\dot a}\,,\\
\comm*{{M_\alpha}^\beta}{Q_{\gamma,a\dot a}}&=\delta_\gamma^\beta Q_{\alpha,a\dot a}-\frac{1}{2}\delta_\alpha^\beta Q_{\gamma,a\dot a}\,,
&\comm*{{M_\alpha}^\beta}{S_{\gamma,a\dot a}}&=\delta_\gamma^\beta S_{\alpha,a\dot a}-\frac{1}{2}\delta_\alpha^\beta S_{\gamma,a\dot a}\,,\\
\comm*{{D}^\beta}{Q_{\alpha,a\dot a}}&=\frac{1}{2}^\beta Q_{\alpha,a\dot a}\,,
&\comm*{{D}^\beta}{S_{\alpha,a\dot a}}&=-\frac{1}{2} S_{\alpha,a\dot a}\,.
\end{align}
For the odd-odd part we get
\begin{align}
\acomm*{Q_{\alpha,a\dot a}}{{Q^\beta}_{b\dot b}}&=\epsilon_{a\dot a}\epsilon_{b\dot b}{P_{\alpha}}^{\beta}\,,\qquad
\acomm*{S_{\alpha,a\dot a}}{{S^\beta}_{b\dot b}}=\epsilon_{a\dot a}\epsilon_{b\dot b}{K_{\alpha}}^{\beta}\,,\\
\acomm*{Q_{\alpha,a\dot a}}{{S^\beta}_{b\dot b}}&=\epsilon_{a\dot a}\epsilon_{b\dot b}\left({M_\alpha}^{\beta}+\delta_\alpha^\beta D\right)-\delta_\alpha^\beta\left(\epsilon_{ab}\bar R_{\dot a\dot b}+\epsilon_{\dot a\dot b}R_{ab}\right)\,.
\end{align}
where $R_{ab}=\epsilon_{bc}{R_a}^c$ and $\bar R_{\dot a\dot b}=\epsilon_{\dot b\dot c}{R_{\dot a}}^{\dot c}$.

We are now in business to give up conformal invariance. We define the spacetime generators 
\begin{equation}
{J^{(\ell)}_\alpha}^\beta=\frac{r}{2}\left(\frac{1}{r}{M_\alpha}^{\beta}-{P_\alpha}^{\beta}-\frac{1}{4r^2}{K_\alpha}^{\beta}\right)\,,\quad
{J^{(r)}_\alpha}^\beta=\frac{r}{2}\left(\frac{1}{r}{M_\alpha}^{\beta}+{P_\alpha}^{\beta}+\frac{1}{4r^2}{K_\alpha}^{\beta}\right)\,,
\end{equation}
which satisfy the algebra
\begin{equation}
\comm*{{J^{(\ell/r)}_\alpha}^\beta}{{J^{(\ell/r)}_\gamma}^\delta}=\delta_\gamma^\beta{J^{(\ell/r)}_\alpha}^\delta-\delta_\alpha^\delta{J^{(\ell/r)}_\beta}^\gamma\,,\qquad
\comm*{{J^{(\ell)}_\alpha}^\beta}{{J^{(r)}_\gamma}^\delta}=0\,.
\end{equation}
If we map our theory on $S^3$, ${J^{(\ell)}_\alpha}^\beta$ and ${J^{(r)}_\alpha}^\beta$ are the generators of the isometry algebra on $S^3$ $\mathfrak{su}(2)_{\ell}$, $\mathfrak{su}(2)_{r}$, respectively. We can interpret $\mathfrak{su}(2)_\ell\oplus\mathfrak{su}(2)_r$ as the bosonic part of  the $\mathfrak{su}(2|1)_\ell\oplus\mathfrak{su}(2|1)_r$, which is an $\mathcal{N}=4$ Poincar\'e superalgebra on $S^3$. More precisely, if we define
\begin{align}
{J^{(\ell/r)}_\alpha}^\beta=\begin{pmatrix}J_3^{(\ell/r)}&J_1^{(\ell/r)}+iJ_2^{(\ell/r)}\\J_1^{(\ell/r)}-iJ_2^{(\ell/r)}& -J_3^{(\ell/r)} \end{pmatrix}\,,
\end{align} 
where $J_i^{(\ell/r)}$ closes the algebra
\begin{equation}
\comm*{J_i^{(\ell/r)}}{J_j^{(\ell/r)}}=i\epsilon_{ijk}J_k^{(\ell/r)}\,,
\end{equation}
and their action on a generic operator $O$ is 
\begin{equation}
\comm*{J_i^{(\ell)}}{O}=\mathcal{L}_{v_i^\ell}O\,, \qquad
\comm*{J_i^{(r)}}{O}=\mathcal{L}_{v_i^r}O\,,
\end{equation}
where $\mathcal{L}_{v_i^\ell}$ ($\mathcal{L}_{v_i^r}$) is the Lie derivative w.r.t. by the left (right) invariant vectors fields of $S^3$
\begin{subequations}\label{s3action}
\begin{align}
v_1^\ell&=\frac{i}{2}\left(-\cos(\tau+\varphi)\partial_\theta-\tan(\theta)\sin(\tau+\varphi)\partial_\tau+\cot(\theta)\sin(\tau+\varphi)\partial_\varphi   \right)   \,,\\
v_2^\ell&=\frac{i}{2}\left(\sin(\tau+\varphi)\partial_\theta-\tan(\theta)\cos(\tau+\varphi)\partial_\tau+\cot(\theta)\cos(\tau+\varphi)\partial_\varphi\right)   \,,\\
v_3^\ell&=\frac{i}{2}\left(\partial_\tau+\partial_\varphi\right)   \,,\\
v_1^r&=\frac{i}{2} \left(\cos(\tau-\varphi)\partial_\theta+\tan(\theta)\sin(\tau-\varphi)\partial_\tau+\cot(\theta)\sin(\tau-\varphi)\partial_\varphi   \right)   \,,\\
v_2^r&=\frac{i}{2} \left(-\sin(\tau-\varphi)\partial_\theta-\tan(\theta)\cos(\tau-\varphi)\partial_\tau+\cot(\theta)\cos(\tau-\varphi)\partial_\varphi\right) \,,\\
v_3^r&=\frac{i}{2} \left(\partial_\tau-\partial_\varphi\right)   \,.
\end{align}
\end{subequations}

The R-symmetry of the $\mathfrak{su}(2)_{\ell}$, $\mathfrak{su}(2)_{r}$ algebra is specified by two matrices ${h_a}^b\in\mathfrak{su}(2)_H$ and ${\bar h_{\dot a}}^{\;\;\dot b}\in\mathfrak{su}(2)_C$, which select a Cartan of the $\mathcal{N}=4$ R-symmetry group as $U(1)_H\times U(1)_C \subset SU(2)_H\times SU(2)_C $
\begin{equation}
R_H=\frac{1}{2}{h_a}^b{R_b}^a\,,\qquad R_C=\frac{1}{2}{\bar h_{\dot a}}^{\;\;\dot b}{\bar{R}_{\dot b}}^{\;\;\dot a}\,.
\end{equation}

As in \cite{Dedushenko:2016jxl}, to construct the supercharges of $\mathfrak{su}(2)_{\ell}$, $\mathfrak{su}(2)_{r}$ in full generality, we decompose ${h_a}^b\in\mathfrak{su}(2)_H$ and ${\bar h_{\dot a}}^{\;\;\dot b}\in\mathfrak{su}(2)_C$ using four vectors $u_{\pm}^a$, $\bar u_\pm^{\dot a}$ in the R-symmetry space, defined by the following conditions
\begin{align}
u_+^au_{-a}= \bar{u}_+^{\dot a}\bar{u}_{-\dot{a}}=1\,, \quad
{h_a}^b=u_{+a}u_-^{b}+u_{-a}u_+^{b}\,, \quad  {\bar h_{\dot a}}^{\;\;\dot b}=\bar u_{+\dot a} \bar  u_-^{\dot b}+ \bar u_{-\dot a} \bar u_+^{\dot b}\,.
\end{align}
Together, they imply that
\begin{equation}
u^a_\pm {h_a}^b=\pm u^b_\pm \,, \qquad  \bar u^{\dot a}_\pm{\bar h_{\dot a}}^{\;\;\dot b}=\pm \bar u^{\dot b}_\pm \,,
\end{equation}
Then, our supercharges are given by
\begin{equation}
Q^{(\ell\pm)}_\alpha=\frac{1+i}{2}u_{\pm}^a\bar u^{\dot a}_\mp\left(Q_{\alpha,a\dot a}-\frac{1}{2r}S_{\alpha,a\dot a}\right)\,,\quad
Q^{(r\pm)}_\alpha=\frac{1+i}{2}u_{\mp}^a\bar u^{\dot a}_\mp\left(Q_{\alpha,a\dot a}+\frac{1}{2r}S_{\alpha,a\dot a}\right)\,.
\end{equation}
We find the following algebra
\begin{align}
\acomm*{Q^{(\ell+)}_\alpha}{Q^{(\ell-)}_\beta}&=\frac{i}{r}\left[J_{\alpha\beta}^{(\ell)}+\frac{1}{2}\epsilon_{\alpha\beta}R_\ell\right] \,, \\
\acomm*{Q^{(r+)}_\alpha}{Q^{(r-)}_\beta}&=\frac{i}{r}\left[J_{\alpha\beta}^{(r)}+\frac{1}{2}\epsilon_{\alpha\beta}R_r\right]\,,  
\end{align}
where we have defined
\begin{equation}
R_\ell=R_C-R_H\,, \qquad R_r=R_C+R_H\,.
\end{equation}
The even/odd commutators are
\begin{align}
\comm*{R_\ell}{Q_\alpha^{(\ell\pm)}}&=\pm Q_\alpha^{(\ell\pm)}\,, &\comm*{R_r}{Q^{(r\pm)}_\alpha}&=\pm Q^{(r\pm)}_\alpha\,, \\
\comm*{{{J^{(\ell)}_\alpha}^\beta}}{Q^{(\ell\pm)}_\gamma}&=\delta_\gamma^\beta Q^{(r\pm)}_\alpha-\frac{1}{2}\delta_\alpha^\beta Q^{(r\pm)}_\gamma  \,,
&\comm*{{{J^{(r)}_\alpha}^\beta}}{Q^{(\ell\pm)}_\gamma}&=\delta_\gamma^\beta Q^{(r\pm)}_\alpha-\frac{1}{2}\delta_\alpha^\beta Q^{(r\pm)}_\gamma\,.
\end{align}
All the other commutators are vanishing. 
Finally, to recover the choice ${h_a}^b=-{(\sigma_2)_a}^b$, ${\bar h_{\dot a}}^{\;\;\dot b}=-{(\sigma_3)_{\dot a}}^{\;\;\dot b}$ we choose, like in \cite{Dedushenko:2016jxl}, the following R-symmetry vectors
\begin{equation}
u^a_+=\frac{1}{2}\begin{pmatrix} 1-i\\1+i\end{pmatrix}\,, \quad
u^a_-=\frac{1}{2}\begin{pmatrix} 1-i\\-1-i\end{pmatrix}\,, \quad
\bar u^{\dot a}_+=\begin{pmatrix} 0\\1\end{pmatrix}\,, \quad
\bar u^{\dot a}_-=\begin{pmatrix} 1\\0\end{pmatrix}\,.
\end{equation}

These $\mathcal{N}=4$ algebras are realized by the transformations which leaves invariant the actions introduced in Sec.~\ref{sec1.1}. The supersymmetry transformations for the vector multiplets are
\begin{subequations}\label{susyN4}
\begin{align}
\delta A_\mu&=\frac{i}{2}\xi^{a\dot{b}}\gamma_\mu\lambda_{a\dot{b}}\,, \\
\delta\lambda_{a\dot{b}}&=-\frac{i}{2}\epsilon^{\mu\nu\rho}\gamma_{\rho}\xi_{a\dot{b}}F_{\mu\nu}-{D_{a}}^c\xi_{c\dot{b}}-i\gamma^\mu{\xi_a}^{\dot{c}}\mathcal{D} _\mu\Phi_{\dot{c}\dot{b}}+2i{\Phi_{\dot{b}}}^{\dot{c}}\xi'_{a\dot{c}}+ \\
&+\frac{i}{2}\xi_{a\dot{d}}\comm*{{\Phi_{\dot{b}}}^{\dot{c}}}{{\Phi_{\dot{c}}}^{\dot{d}}} \,,    \notag   \\ 
\delta\Phi_{\dot{a}\dot{b}}&={\xi^c}_{(\dot{a}}\lambda_{|c|\dot{b})} \,, \\
\delta D_{ab}&=-i\mathcal{D}_\mu ({\xi_{(a}}^{\dot{c}}\gamma^\mu\lambda_{b)\dot{c}})-2i{\xi'_{(a}}^{\dot{c}}\lambda_{b)\dot{c}}+i\comm*{{\xi_{(a}}^{\dot{c}}\lambda_{b)}^{\dot{d}}}{\Phi_{\dot{c}\dot{d}}}  \,.
\end{align}
\end{subequations}
For the hypermultiplet, we have
\begin{subequations}\label{susyh}
\begin{align}
\delta q^a&=\xi^{a\dot{b}}\psi_{\dot{b}}\,,  &\delta_\xi\psi_{\dot{a}}&=i\gamma_\mu\xi_{a\dot{a}}\mathcal{D} _\mu q^a+i\xi'_{a\dot{a}}q^a-i\xi_{a\dot{c}}\Phi\indices{^{\dot{c}}_{\dot{a}}}q^a \,, \\
\delta\tilde{q}^a&=\xi^{a\dot{b}}\tilde{\psi}_{\dot{b}}\,,         &\delta_\xi \tilde{\psi}_{\dot{a}}&=i\gamma_\mu\xi_{a\dot{a}}\mathcal{D} _\mu \tilde{q}^a+i\xi'_{a\dot{a}}\tilde{q}^a-i\xi_{a\dot{c}}\Phi\indices{^{\dot{c}}_{\dot{a}}}\tilde{q}^a  \,.
\end{align}
\end{subequations}
We take $\xi_{a\dot b}$ to be conformal Killing spinors
\begin{equation}
\nabla_\mu\xi_{a\dot a}=\gamma_\mu\xi'_{a\dot a}\,, \qquad
\nabla_\mu\xi'_{a\dot a}=-\frac{1}{4r^2}\gamma_\mu\xi'_{a\dot a}\,.
\end{equation}
On $S^3$, in stereographic coordinates $x^i$ with $i=1,2,3$ and in the stereographic frame \footnote{ The stereographic frame is defined as $e^i_\mu=e^\Omega\,\delta_\mu^i$, with $$e^\Omega=\frac{1}{1+\frac{x_i^2}{4r^2}}=\frac{1+\sin\theta\cos\varphi}{2}\,,$$ being the conformal factor. We refer to the App.~A of \cite{Dedushenko:2016jxl} for details)}, the explicit solution is
\begin{subequations}
\begin{align}
\xi_{a\dot a}&=e^{\Omega/2}\left(\theta_{a\dot a}-x^i\gamma_i\eta_{a\dot a}\right)\,,\\
\xi_{a\dot a}&=e^{\Omega/2}\left(\eta_{a\dot a}+\frac{1}{4r^2}x^i\gamma_i\theta_{a\dot a}\right)\,.
\end{align}
\end{subequations}
In the flat space limit $r\to\infty$, we recover the familiar form of flat space conformal Killing spinor of Eq.~\eqref{flatks}
\begin{equation}
\xi_{a\dot a}=\theta_{a\dot a}-x_i\gamma_i \eta_{a\dot a}\,, \qquad \xi'_{a\dot a}=-\eta_{a\dot a}\,.
\end{equation}
The Killing spinors associated with the $\mathfrak{su}(2|1)_\ell\oplus\mathfrak{su}(2|1)_r$ superalgebra on $S^3$ satisfy the additional constraint
\begin{equation}
\xi_{a\dot a}=-\frac{i}{2r}{h_a}^b\,\xi'_{b\dot b}\,\bar{h}_{\dot a}^{\;\;\dot b}\,.
\end{equation}
We can recover the explicit action of the bosonic generators of \eqref{s3action} from the following definition 
\begin{equation}
\delta\Phi=\comm*{\theta^{a\dot a }Q_{a\dot a}-i\eta^{a\dot a} S_{a\dot a}}{\Phi}\,.
\end{equation}

\bibliography{biblio}

\end{document}